\documentclass{mn2e}
\usepackage[dvips]{graphicx}
\usepackage{amssymb}

\begin{document}

  \title[Formation and survival of discs in $\Lambda$CDM]{The formation and survival of discs in a $\Lambda$CDM universe}

\author[Scannapieco et al.]{Cecilia Scannapieco $^{1}$\thanks{E-mail: cecilia@mpa-garching.mpg.de},
Simon D.M White$^{1}$, Volker Springel$^{1}$ and  Patricia B. Tissera$^{2,3}$
\\
$^1$ Max-Planck Institute for Astrophysics, Karl-Schwarzchild Str. 1, D85748, Garching, Germany.\\
$^2$ Instituto de Astronom\'{\i}a y F\'{\i}sica del Espacio, Casilla de Correos 67,
Suc. 28, 1428, Buenos Aires, Argentina.\\
$^3$  Consejo Nacional de Investigaciones Cient\'{\i}ficas
y T\'ecnicas, CONICET, Argentina.\\ 
}

   \maketitle

   \begin{abstract}

     We study the formation of galaxies 
in a $\Lambda$ cold dark matter ($\Lambda$CDM) universe using high resolution  hydrodynamical simulations
with a multiphase treatment of gas, cooling and feedback, 
focusing on the formation of discs. 
Our simulations follow eight  isolated
haloes similar in mass to the Milky Way and 
extracted from a large cosmological simulation without
restriction on spin parameter or merger history. 
This  allows us  to  investigate  how the
final properties of the simulated galaxies correlate with the formation
histories of their haloes.
We find that, at $z=0$, 
none of our galaxies contain a disc with more than $20$ per cent of
its total stellar mass. Four of the eight  galaxies nevertheless have well-formed disc components, 
three have dominant spheroids and very small discs, and 
one is a spheroidal galaxy with no disc at all. 
The $z=0$ spheroids are made of old stars, while discs are younger and
formed from the inside-out.
 Neither the existence of a disc at $z=0$ 
nor the final disc-to-total mass ratio seems to depend on the spin parameter
of the halo. Discs are formed in haloes with spin parameters
as low as $0.01$ and as high as $0.05$;  galaxies
with little or no disc component span the same range in
spin parameter.
Except for one of the simulated galaxies, all
have significant discs at $z\gtrsim 2$, 
regardless of their $z=0$ morphologies. Major mergers and
instabilities, which arise when accreting cold gas is misaligned
with the  stellar disc,
 trigger a transfer of mass from the discs
to the spheroids. In some cases, discs are destroyed, while
in others, they survive or reform. This suggests that the survival
probability of discs depends on the particular formation history
of each galaxy. A realistic $\Lambda$CDM model will clearly require
weaker star formation at high redshift and later disc assembly
than occurs in our models.

   \end{abstract}


\section{Introduction}

According to the standard picture of cosmic
structure formation, dark matter and baryons 
acquire similar specific angular momentum  during their
collapse through external torques from  neighbouring
structures (Hoyle 1953; Peebles 1969; White 1984).
Discs are formed as the gas cools and condenses within the
dark matter haloes conserving its specific angular
momentum
(White \& Rees 1978; Fall \& Efstathiou 1980; Mo, Mao \& White 1998).
In these models, the final properties of discs are
 closely linked
to those of their parent haloes:
for example,  disc sizes and stability
properties depend on the spin parameter of the dark matter
haloes in which they reside.

However, discs are  unstable against rapid changes
in the gravitational potential, such as those
produced by the accretion of satellites 
(Toth \& Ostriker 1992; Quinn, Hernquist \& Fullagar 1993;
Velazquez \& White 1999). 
Mergers of gas-poor disc galaxies 
are expected to produce  early-type galaxies
as remnants (Toomre \& Toomre 1972; Toomre 1977).
Isolated galaxy mergers have been extensively studied
with hydrodynamical simulations, confirming that major
mergers can indeed transform disc galaxies into elliptical-like 
systems
(Barnes 1992; Barnes \& Hernquist 1992, 1996; Mihos \& Hernquist 1994, 1996). 
Given the fragility of discs and
the fact that hierarchical models predict
mergers to occur quite frequently, the
existence of present-day spiral galaxies
has been interpreted as an indirect proof of late
disc formation in systems with a quiescent recent past.

Numerical hydrodynamical simulations are a powerful tool to study galaxy
formation and evolution in its full cosmological context, since mergers are 
then naturally accounted for. 
State-of-the-art simulations currently include
treatments of star formation, cooling, chemical enrichment
and supernova (SN) feedback, and in some cases other processes
such as black hole formation, active galactic nuclei (AGN) feedback and/or cosmic
ray generation. These have been  successful in reproducing
some observed properties of galaxies, such as
overall scaling relations,
star formation histories and chemical abundances
(e.g. 
Springel \&
Hernquist 2003; Brook et al. 2004; Okamoto et al. 2005; Oppenheimer \& Dav\'e 2006;
Stinson et al. 2006; Governato et al. 2007;
Dalla Vecchia \& Schaye 2008; Dubois \& Teyssier 2008; Finlator \& Dav\'e 2008;
Scannapieco et al. 2008 (hereafter S08); 
among recent examples).
However, 
  such simulations have generally had difficulty reproducing disc-dominated galaxies
in typical dark matter haloes, when taking into account  the cosmological
setting. A major problem is known as the ``angular momentum catastrophe''.
This occurs when baryons condense early and then transfer
a significant fraction of their angular momentum to the
dark matter as the final galaxy is assembled
(Navarro \& Benz 1991; Navarro \& White 1994).
As a result, discs contain too small a fraction of the stellar mass
in comparison to  observed spirals.

In a limited number of  cases, 
 simulations have been able to produce
individual examples of
realistic disc galaxies in $\Lambda$ cold dark matter ($\Lambda$CDM)  universes
(e.g. Abadi et al. 2003; Governato et al. 2004;
Robertson et al. 2004; Okamoto et al. 2005; Governato et al. 2007;
S08; see also Croft et al. 2008).
Though, none has produced a viable late-type spiral. 
These simulations use the  `zoom' technique
which allows  an individual galaxy to be simulated at high
resolution in its proper cosmological context.
The relative success of these models appears to reflect
the inclusion of efficient supernova feedback
and improved numerical resolution.
Typically, the simulated haloes
are chosen so as to improve the chances of disc formation:
  masses similar to the
Milky Way, relatively high spin parameters, and quiet merger histories.
Such haloes are believed to be the most likely to produce disc galaxies, 
but given our incomplete knowledge of galaxy formation
and evolution, and our repeated failure to
reproduce late-type spirals, it is  uncertain whether these
assumptions are correct.

In a recent paper (S08), we 
studied the effects of SN  feedback on the
formation of galaxy discs by using cosmological
simulations of a Milky Way-type galaxy. 
This study  used an extension of the
Tree-PM SPH code {\small GADGET-2} (Springel 2005)
which includes star formation, chemical enrichment and
SN feedback, metal-dependent cooling and a multiphase
model for the gas component (Scannapieco et al. 2005, 2006; hereafter
S05, S06 respectively).
Unlike other schemes, this treatment does not require ad hoc elements
such as suppression of cooling  after SN explosions or
the explicit insertion of winds. Galactic
winds are {\it naturally} generated when the multiphase
gas model and SN feedback are included.
In S08, we  showed that this model can
produce realistic discs from cosmological
initial conditions. This is a consequence of self-regulation of
the star formation and the generation of galactic winds.
Most combinations of the input
parameters yield  disc-like
components, although with different sizes and thicknesses, 
indicating that the code
can form discs without fine-tuning the implemented physics. 
In all cases tested, however, the dominant stellar component by
mass was the spheroid.

In this paper, we investigate these issues further by following
the evolution of 
eight different  galaxy-halo systems in a $\Lambda$CDM universe at
higher resolution than in our previous work. We use 
an updated simulation code {\small GADGET-3} (Springel
et al. 2008)  which includes our multiphase
gas and feedback algorithms.
Target haloes, 
similar in mass as the Milky Way and mildly isolated,  were 
chosen  from a  dark matter only simulation of a cosmological
volume, with  no
restriction on their spin parameter or merger history.
These haloes were resimulated including baryons and at improved
mass and force resolution. 
Our main goal here  is to
test whether realistic discs form
in typical $\Lambda$CDM haloes and whether  spin parameter or merger history
influences their properties
as usually assumed.

Our paper is organized as follows. In Section~\ref{sims},
we describe the main characteristics of the initial conditions
and the implemented physics. In Section~\ref{results},
we present the  properties of our simulated galaxies
at $z=0$, while in Section~\ref{halo_prop},
we investigate the relation between the present-day
morphology of the galaxies and the properties
of their parent dark matter haloes. In Section~\ref{disk_survival},
we study the evolution of discs with time
and investigate the processes leading to a morphological
transformation. Finally, in Section~\ref{conclu}, we summarize
our main results.

\section{Methodology}

\label{sims}

In this section, we describe
the initial conditions  used for our study,
as well as  the simulation code
and the implemented physics. We also summarize 
previous results which led us to the
 input parameters assumed in our simulations.

\subsection{Initial conditions and simulation setup}

In this paper, we study a set of
 eight galaxy-sized haloes which
were selected as part of the Aquarius Project of the Virgo
Consortium (Springel et al. 2008).
Target haloes were chosen at $z=0$  from a simulation of a cosmological box 
of $137$ Mpc on a side. New initial conditions
were then constructed with improved mass resolution.
The initial conditions  originally  had  dark matter
particles only because the main Aquarius Project studies the properties
of the dark matter distribution.  
Gas particles were added to these initial conditions for our simulations,
displaced by half the interparticle separation,
and  the mass of dark matter particles was reduced accordingly.
We refer to Springel et al. (2008) for a detailed
explanation of the generation of the initial conditions.
Target haloes were chosen to have similar
mass as the Milky Way and 
 to be
mildly isolated at $z=0$ 
(no neighbour exceeding half
of their mass within $1.4$ Mpc).
Except for these
conditions, the haloes were selected   randomly
from the parent simulation.

Our simulations follow the evolution of matter
from $z=127$ to $z=0$ 
in a $\Lambda$CDM universe with the following cosmological parameters: 
$\Omega_\Lambda=0.75$, $\Omega_{\rm m}=0.25$,
$\Omega_{\rm b}=0.04$,  
$\sigma_8=0.9$ and 
$H_0=73$ km s$^{-1}$ Mpc$^{-1}$.
The masses of dark matter and gas particles used in 
each simulation are listed
in Table~\ref{simulations_table}.
We have adopted 
 the same  gravitational 
softening for dark matter, gas and star particles, which
varies in the range $0.7-1.4$ kpc for the different
simulations\footnote{Different simulations have 
been run with slightly different softenings.
We have, however, performed a number of experiments to test the
effects of varying the adopted softening within this range, 
and found that the results are not sensitive to this choice.}.

The simulated galaxies have  $z=0$  virial masses in the range
$\sim 7-16\times 10^{11}$ 
M$_\odot$ and are represented by about
$1$ million particles within their virial radius.
At $z=0$, baryon fractions are about
$0.10$ within this radius, which are
smaller than the global value of $0.16$, indicating 
that  a significant amount of baryons has
been lost through winds.
The main characteristics of the haloes, at $z=0$,
are listed in Table~\ref{simulations_table}.
We show the virial radius ($r_{200}$) and virial mass ($M_{200}$), 
the masses in stars and in gas within the virial radius, 
the optical radius ($r_{\rm opt}$, 
defined to enclose $83$ per cent of
the cold gas plus stellar mass), the baryon fraction within $r_{200}$, and the spin parameter, calculated
using Eq.~(5) of Bullock et al. (2001) at $r_{200}$.
Following the convention of Springel et al. (2008)
for halo naming,
each of our haloes has a reference letter (``A" to ``H'')
and a number ($5$ for all our simulations)
which denotes the resolution level.
Our sample includes the six haloes studied in 
Springel et al. (2008), 
and two additional ones. All of them
were selected according to the same criteria.

In Table~\ref{simulations_table}, we also show 
results for two lower resolution versions of simulation
Aq-E-5. We find that the three Aq-E simulations give
convergent results within a few per cent for the global
properties of the galaxies (virial radius, virial mass, spin parameter).
However, larger differences are detected for the final stellar mass,
in particular for Aq-E-7.
The lowest and intermediate resolution runs, at $z=0$, 
respectively formed $45$ and $10$ per cent less stellar mass than the
reference Aq-E-5 simulation. 
These results suggest that
the main set of simulations presented in this paper 
have converged reasonably well (see also Section~\ref{results}.)

\begin{table*} 
\begin{small}
\caption{Principal characteristics of the simulated haloes, at $z=0$:
virial radius ($r_{\rm vir}$), virial mass ($M_{\rm vir}$), 
mass in stars ($M_{\rm stars}$) and gas ($M_{\rm gas}$), optical radius ($r_{\rm opt}$), 
 spin parameter ($\lambda `$)
calculated using Eq.~(5) of Bullock et al. (2001) and baryon fraction. 
We also show the dark matter and gas particle masses.}
\vspace{0.1cm}
\label{simulations_table}
\begin{center}
\begin{tabular}{lccccccccc}
\hline
Galaxy  & $r_{\rm vir}$  & $M_{\rm vir}$ & $M_{\rm star}$&$M_{\rm gas}$ &  $r_{\rm opt}$ &$\lambda '$ &$f_{\rm b}$ &  $m_{\rm DM}$&  $m_{\rm gas,0}$\\
          & [kpc]& [$10^{10}$M$_\odot$] &   [$10^{10}$M$_\odot$] &  [$10^{10}$M$_\odot$] &  [kpc] && &  [$10^{6}$M$_\odot$] & [$10^{6}$M$_\odot$] \\\hline

Aq-A-5   &   231.9   &   149.2 &    9.00 &  4.63 &           17.9 & 0.017 & 0.09 &2.6 & 0.56 \\

Aq-B-5    &  180.5 &      71.1 &    3.96 &  1.67 &       17.7 & 0.031 &  0.08 & 1.5 & 0.29 \\

Aq-C-5    &     237.4 &      160.7 &    10.8&    3.56&   16.0 & 0.012 & 0.09 & 2.2 & 0.41 \\

Aq-D-5&      233.4 &     149.0 &          7.89&   3.21&  14.8 & 0.019 & 0.07 & 2.3 &0.22\\

Aq-E-5   & 205.9 &      107.5&     8.42&   2.61&         10.6 & 0.026 & 0.10 & 1.8 &0.33\\

Aq-F-5 &   195.8  &   91.2    &    7.70 & 1.68       &  14.1 & 0.049 & 0.10 & 1.2 &0.23\\

Aq-G-5     &    179.6   &   68.1&          4.40 & 1.53& 14.1 & 0.048 &0.09 & 1.2 &0.23\\

Aq-H-5     &   181.8&      73.5&        6.48 &  0.52  &  10.4 & 0.008 &0.10 & 1.4 &0.25\\\hline

Aq-E-6   & 206.4 &      106.9&     7.42&   2.71&         11.8 & 0.027 & 0.10 & 7.4 &1.41\\

Aq-E-7   & 200.5 &    99.72  &   4.73  &  2.67 &         12.8 & 0.031  & 0.07 & 20.8  &3.97\\

\hline

\end{tabular}
\end{center}
\end{small}
\end{table*}

\subsection{The simulation code and implemented physics}

The simulations analysed in this paper were run with an
extended version of the code {\small GADGET-3}, an optimized
version of {\small GADGET-2} (Springel
\& Hernquist 2002; Springel 2005). {\small GADGET-3}
is a lagrangian-based code which uses a Tree-PM solver for the gravitational physics
and the Smoothed Particle Hydrodynamics (SPH) technique
for the gasdynamics.
Our model includes a treatment for star formation
and supernova  feedback, as well as metal-dependent
cooling and a multiphase model for the gas component.
Here, we describe the main characteristics of our implementation,
and we refer the interested reader to S05 and S06
for details.
We note that our implementation of  star formation and feedback is
different from that of Springel \& Hernquist (2003), although
we do use their treatment for a UV background,
based on the formulation of Haardt \& Madau (1996).

\subsubsection{Star formation}
We assume that gas particles are eligible to form stars if
they are denser than a threshold density ($\rho_{\rm th} = 7\times 10^{-26}$ 
g cm$^{-3}$) and if they lie in a convergent flow. For these particles,
we assume a star formation rate (SFR) per unit volume equal to
\begin{equation}
\dot\rho_\star = c\,\frac{\rho}{\tau_{\rm dyn}} ,
\end{equation}
where $c$ is the star formation efficiency, $\rho$ and
 $\rho_*$ are the gas and stellar densities, respectively, 
and $\tau_{\rm dyn} = 1 / \sqrt{4\pi G\rho}$
is the local dynamical time of the gas.
The star formation efficiency $c$  sets the efficiency
at which stars are formed or, equivalently, the typical
time-scale of the star formation process.
From gas particles eligible for star formation, 
stars  are created stochastically 
(Springel \& Hernquist 2003), and
a maximum of two star particles are allowed to form
from each gas particle.

\subsubsection{Cooling}
We use  metal-dependent cooling functions (Sutherland \& Dopita 1993)
which take into account the dependence of cooling on the gas metallicity.
We have a series of seven tables which correspond to different metallicities,
from primordial to [Fe/H]$=0.5$, from which we interpolate as needed. 
The inclusion of such metal-dependent cooling functions is
important, since it has been shown that neglecting metals
can lead to significant  overestimation of the gas
cooling times (e.g. S05; Kobayashi, Springel
\& White 2007).

\subsubsection{Chemical enrichment}
We include a treatment for the generation of Type II and Type
Ia SN explosions (SNII and SNIa, respectively).
For both types of events we compute, at each time-step, 
the  number of exploding stars
as well as the corresponding chemical production. 
For SNII, the number of exploding stars is calculated
by convolving a Salpeter initial mass function (with lower and upper
cut-offs of $0.1$ and $40$ M$_\odot$) with
a metal-dependent life-time for massive stars (Raiteri, Villata \& Navarro 1996)
and assuming that SNII are generated by stars  more massive
than $8$ M$_\odot$. For SNIa, we instead assume a rate of $0.3$
relative to SNII (van den Bergh 1991) and a life-time taken randomly
for each star within the interval $\tau_{\rm SNIa}=[0.7,2]$ Gyr.
We use the Woosley \& Weaver (1995) chemical
yields for SNII, which are metal-dependent and vary 
with relative metal abundances, and those of Thielemann, Nomoto \& Hashimoto (1993) for
SNIa.
Chemical elements are distributed to the SPH neighbours of
exploding stars, with a weight given by the smoothing kernel.
Thereafter, there is no further mixing.

\subsubsection{Multiphase gas model}
We include a multiphase model for the gas component in order to avoid the
artificial overcooling common in standard SPH codes (Thacker et al. 2000; Pearce et al. 2001).  
This model {\it decouples}
particles with dissimilar properties, preventing them from being SPH
neighbours. In practice, we {\it decouple} a given particle $j$ from particle $i$,
meaning that $j$ is explicitly excluded from the neighbour list of $i$, if the
following conditions are fulfilled:
\begin{equation}
\label{entropy_condition}
A_i > \alpha A_j \,
\end{equation}
and
\begin{equation}
\label{shock_condition}
\mu_{ij} < c_{ij} \, 
\end{equation}
where $\alpha$ is a constant, $A_i$ is the entropic function of particle $i$
(Springel \& Hernquist 2002),
 $c_{ij}$ is the pair-averaged sound speed,
and $\mu_{ij}$  measures  the relative
velocity of particles $i$ and $j$ 
projected on to their separation vector (Springel 2005). 
 The latter condition  is
included to avoid decoupling in shock waves, which
can lead to unphysical effects when particles on opposite sides of a
shock do not  see each other (Marri \& White 2003).

This multiphase model has been  designed to improve the description of  hot, 
diffuse material in the context of the SPH technique. One
of  its major advantages over previous implementations 
is that no scale-dependent parameters 
are needed to get realistic behaviour and to generate outflowing
winds from a wide variety of strongly star-forming systems. This treatment 
allows the coexistence of diffuse and dense gaseous components,
and can be combined with a more effective scheme for injecting the energy 
and heavy elements from SNe into the distinct components of the  interstellar
medium (S06).

\subsubsection{Feedback}
We assume that each SN injects $E_{\rm SN}\times 10^{51}$ erg of {\it thermal}
energy into the interstellar medium.
At the time of a SN event, we calculate the total amount
of energy produced by the star particle, and distribute
it between its {\it cold} (temperature  $T < T_{\rm crit}=8\times 10^4$ K and density $\rho > 0.1\rho_{\rm th}$)
and {\it hot} (otherwise) gas neighbours.
The fraction distributed into the cold neighbours is our so-called
{\it feedback parameter $\epsilon_c$}. The distribution of energy
works differently for the different gas phases. Hot neighbours
absorb the energy as thermal energy simultaneously with the SN explosions.
Cold neighbours, instead, accumulate energy from successive SNe
into a reservoir,
and effectively receive it as thermal energy only after a time-delay,
 when the total accumulated energy
is sufficient to raise the particle's entropy to a value similar
to its {\it local hot} neighbours (see S06 for details). 
This procedure is coupled to the ejection of metals, which are
distributed  to the cold and hot phases in proportions, also
set by the $\epsilon_c$ value, although  enrichment
always occurs simultaneously with the explosions.

\subsection{Model advantages and input parameters}

Our combined multiphase and SN feedback treatment is among the
most detailed and least ad hoc models to study the formation
of galaxies in a cosmological context so far developed. 
It does not
include  mass-dependent input parameters, and so is suitable
 for tracking the formation of systems
with a wide range of properties 
from  early times, when they were much smaller
than today.  The model adapts
to different mass systems since both the hot and cold gas
properties are locally  determined by cooling, hydrodynamics
and feedback 
at each time without reference to the global properties
of the system.
There is no need to
include the ad hoc elements typically introduced
in other schemes, such as cooling suppression after SN events,
or ``by-hand'' wind generation. Our code, on the contrary, {\it naturally}
generates galactic winds in  galaxies of all masses, with a strength that
reflects the  depth of the corresponding potential well. 
This results in  a unique tool to study
the formation of galaxies in a cosmological context.

There are nevertheless input parameters related to the SN and multiphase
gas models which need to be set.
Recently, S08 carried out a series of
simulations of a Milky Way-type halo varying the input parameters, and 
analysing the effects  on
galactic properties such as morphology,
disc size, and disc and bulge mean ages. 
They found  parameter ranges which produce 
galaxies of size and thickness similar to 
the Milky Way. Provided the input parameters are
within these intervals,  results
are not very sensitive to their exact values.

The simulations presented in this paper were 
run using (almost)  the same input parameters.
We have used a star formation efficiency of $c=0.1$,
a decoupling parameter for the multiphase model
of $\alpha=50$ and a feedback parameter of $\epsilon_c=0.5$.
We  used the same characteristics
for the chemical model (explained above).
The only parameter that was  changed slightly between
 simulations was $E_{\rm SN}$, the energy input per SN, 
which was either $1$ (for Aq-A-5, Aq-B-5 and Aq-D-5) or $0.7$ 
(for Aq-C-5, Aq-E-5, Aq-F-5, Aq-G-5 and Aq-H-5),  
in units of $10^{51}$ erg. Although
this variation might influence slightly 
 the final properties
of the simulated galaxies, it does not affect the overall
results in a significant way.

\section{Properties of discs and spheroids at the present time}
\label{results}

\begin{figure*}
{\includegraphics[width=40mm]{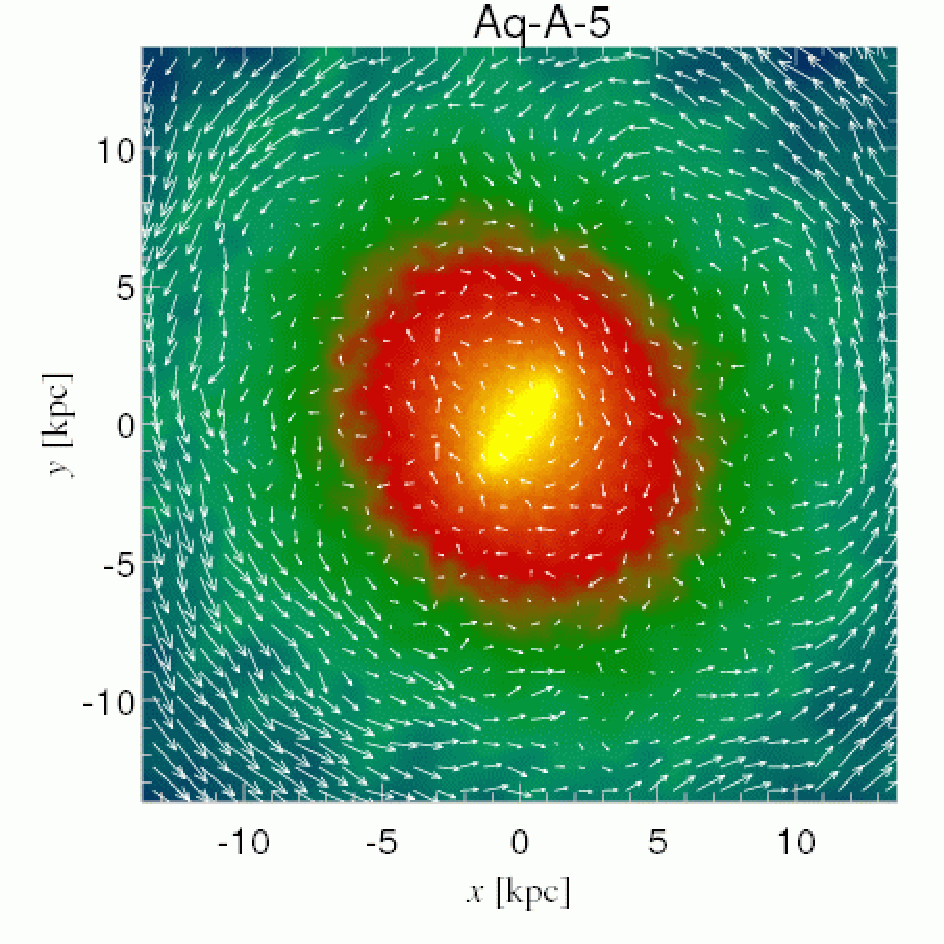}\includegraphics[width=40mm]{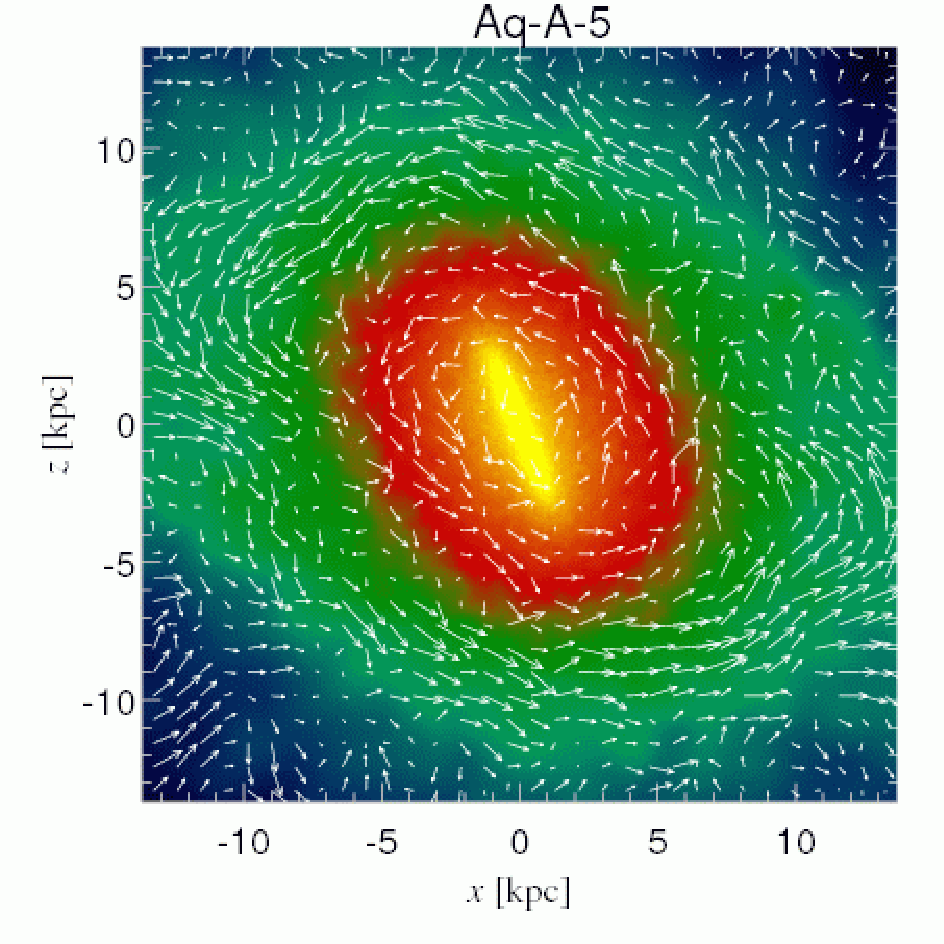}\includegraphics[width=40mm]{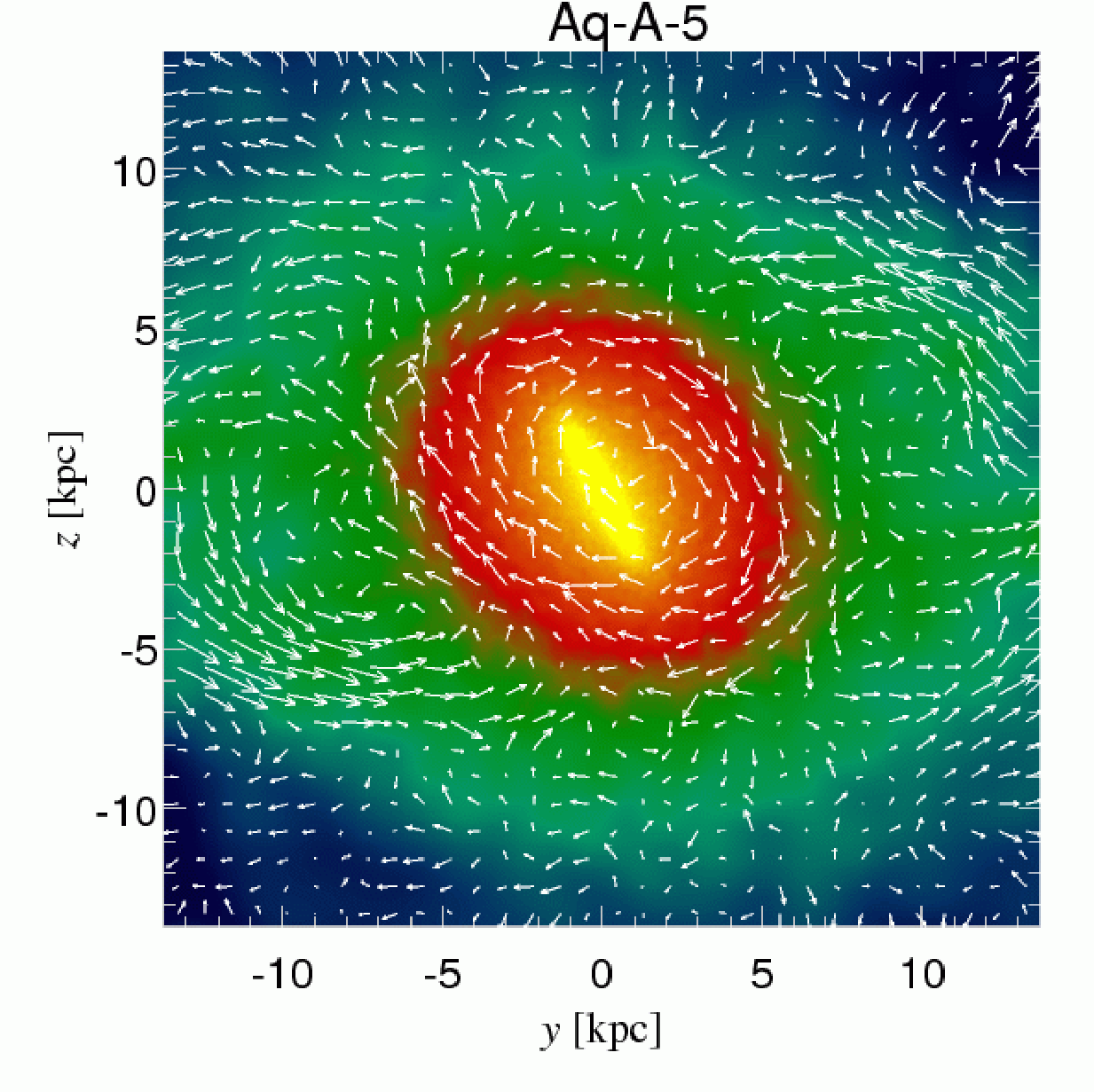}}
{\includegraphics[width=40mm]{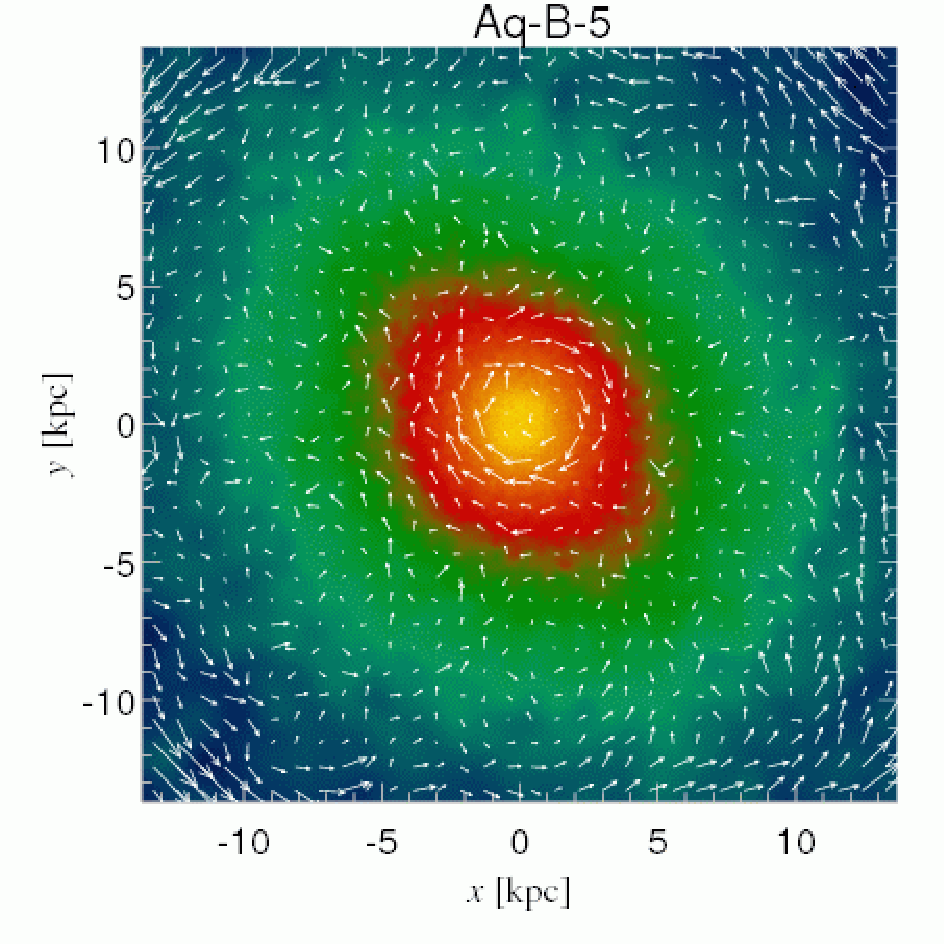}\includegraphics[width=40mm]{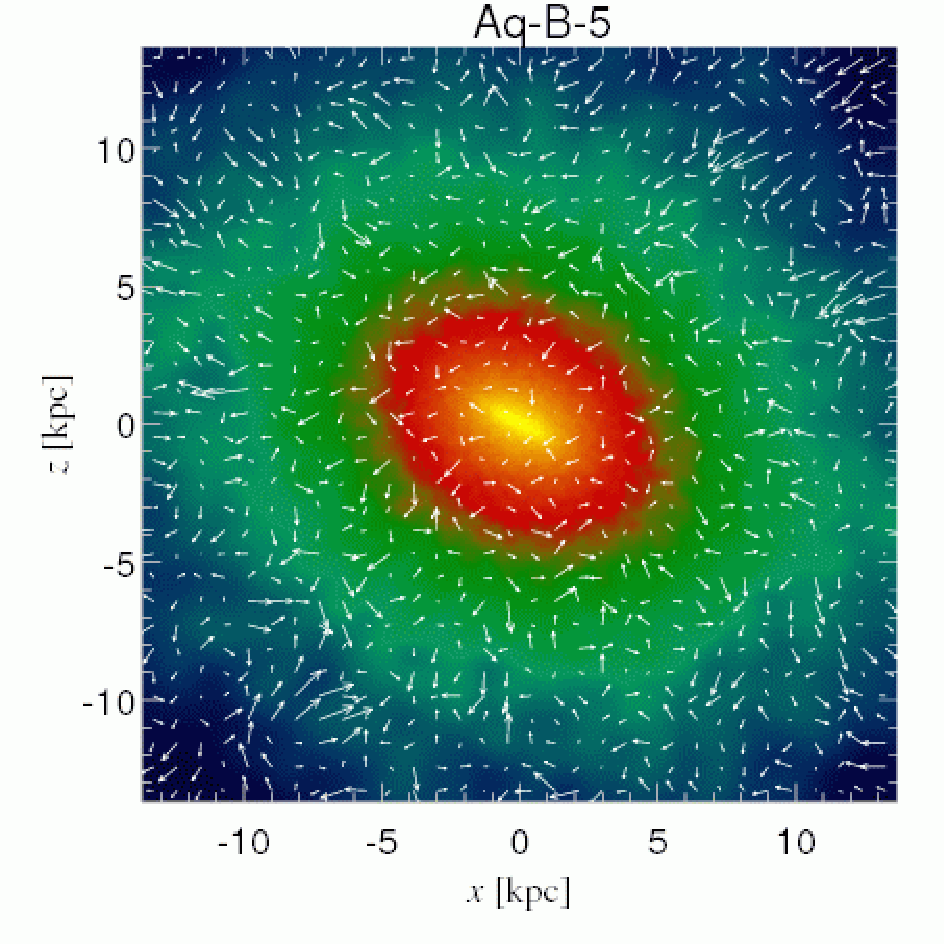}\includegraphics[width=40mm]{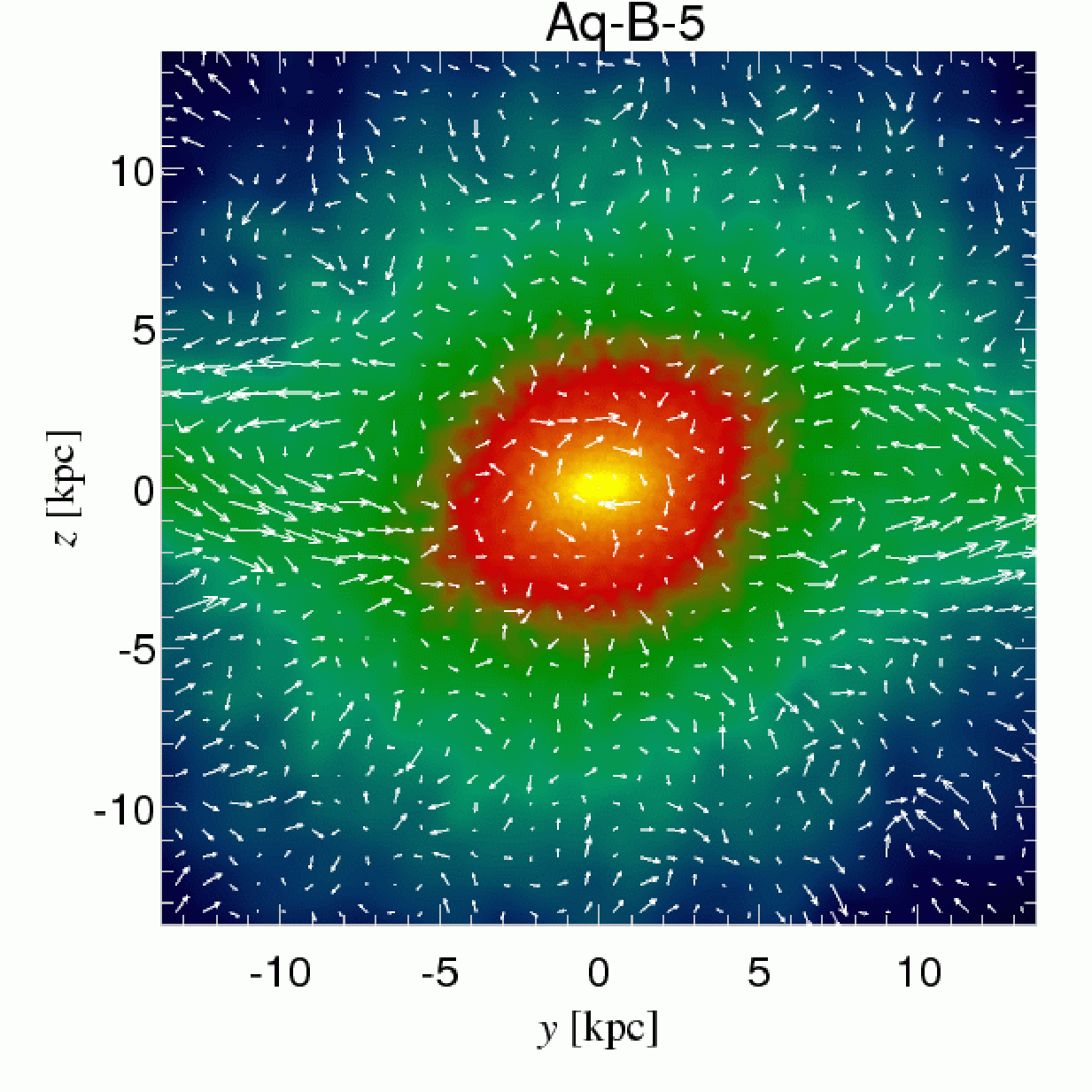}}
{\includegraphics[width=40mm]{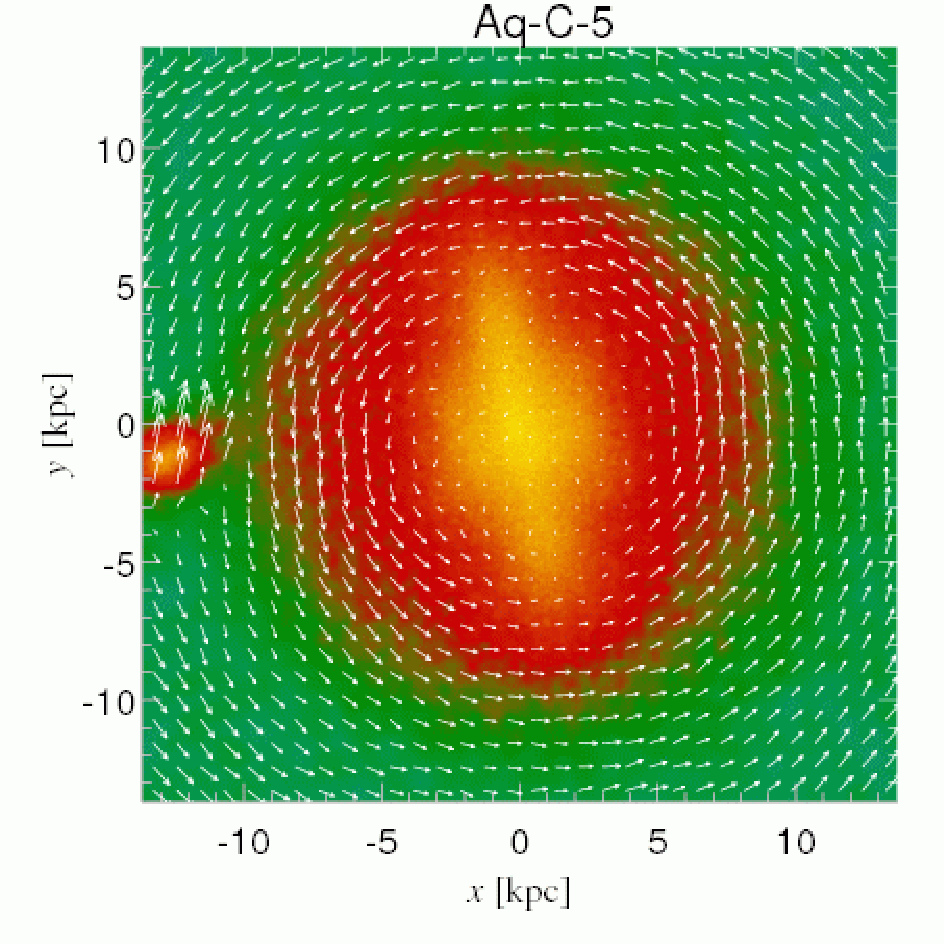}\includegraphics[width=40mm]{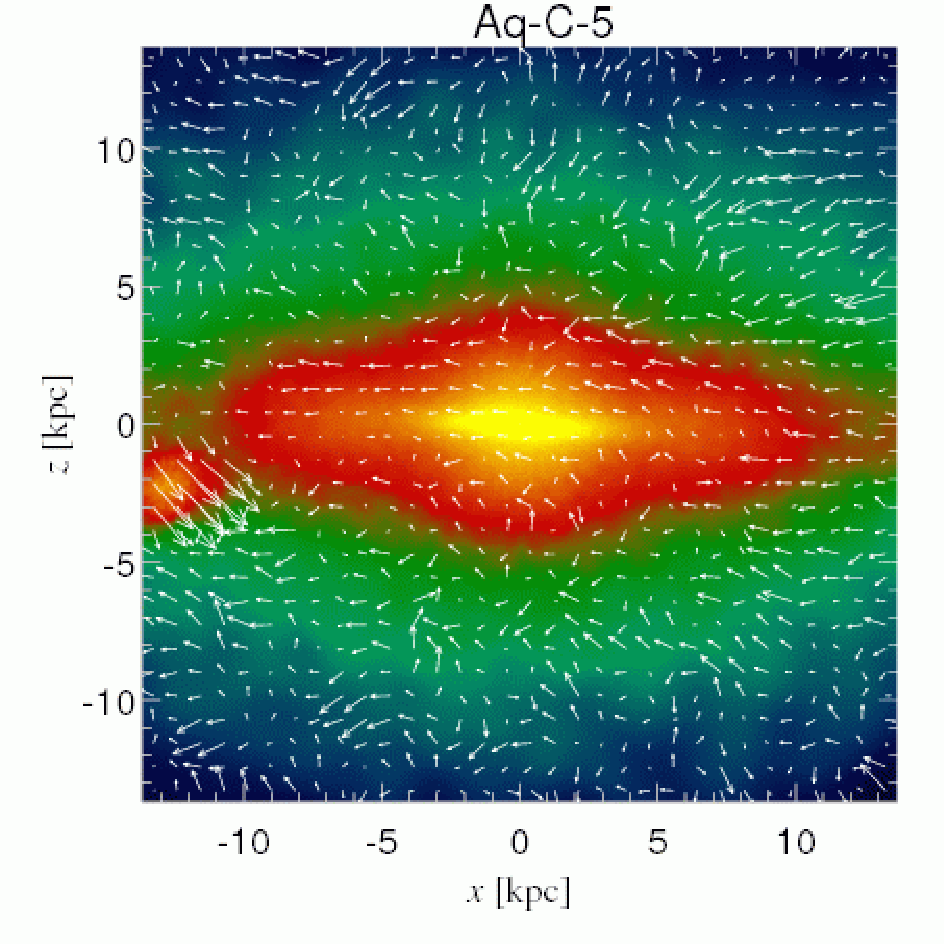}\includegraphics[width=40mm]{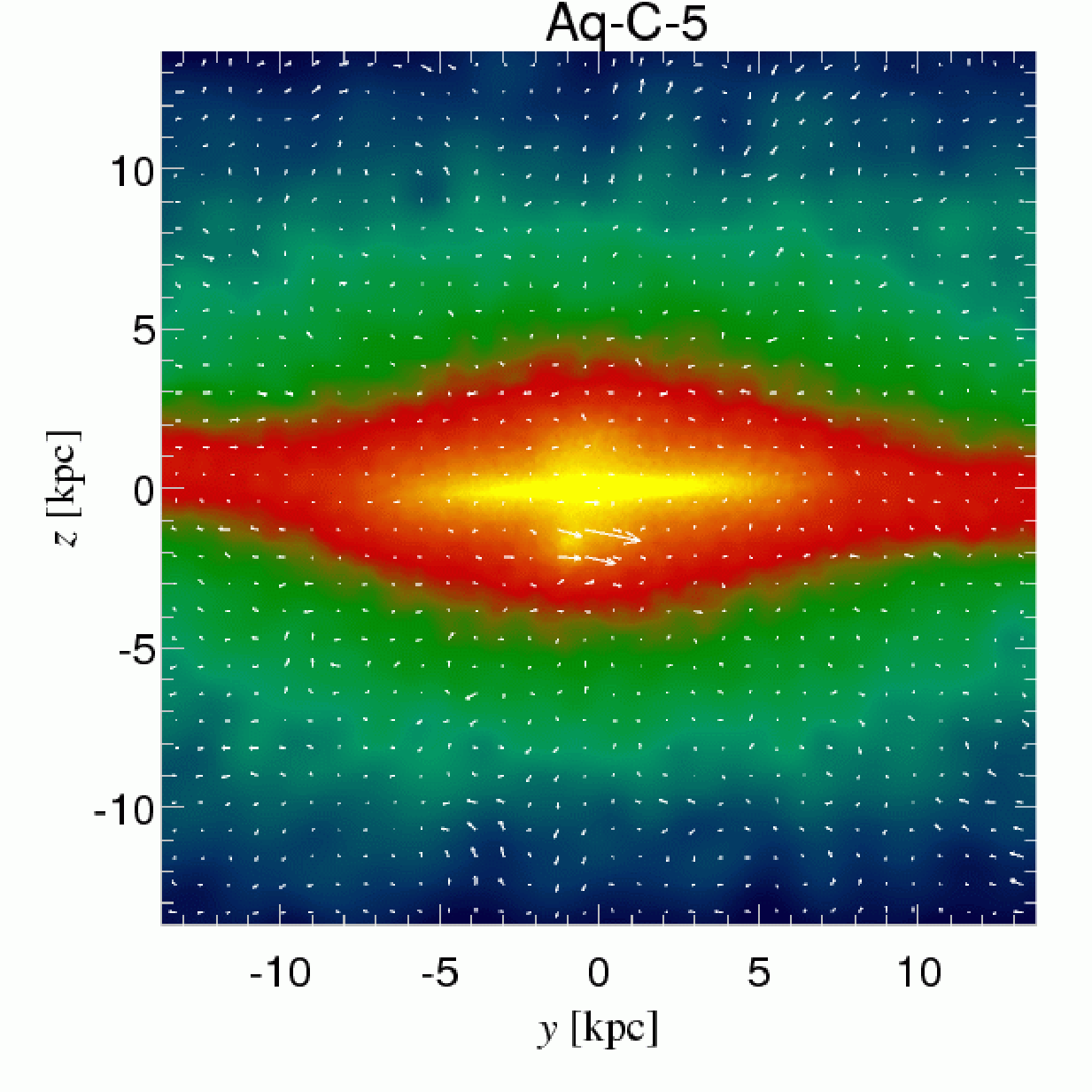}}
{\includegraphics[width=40mm]{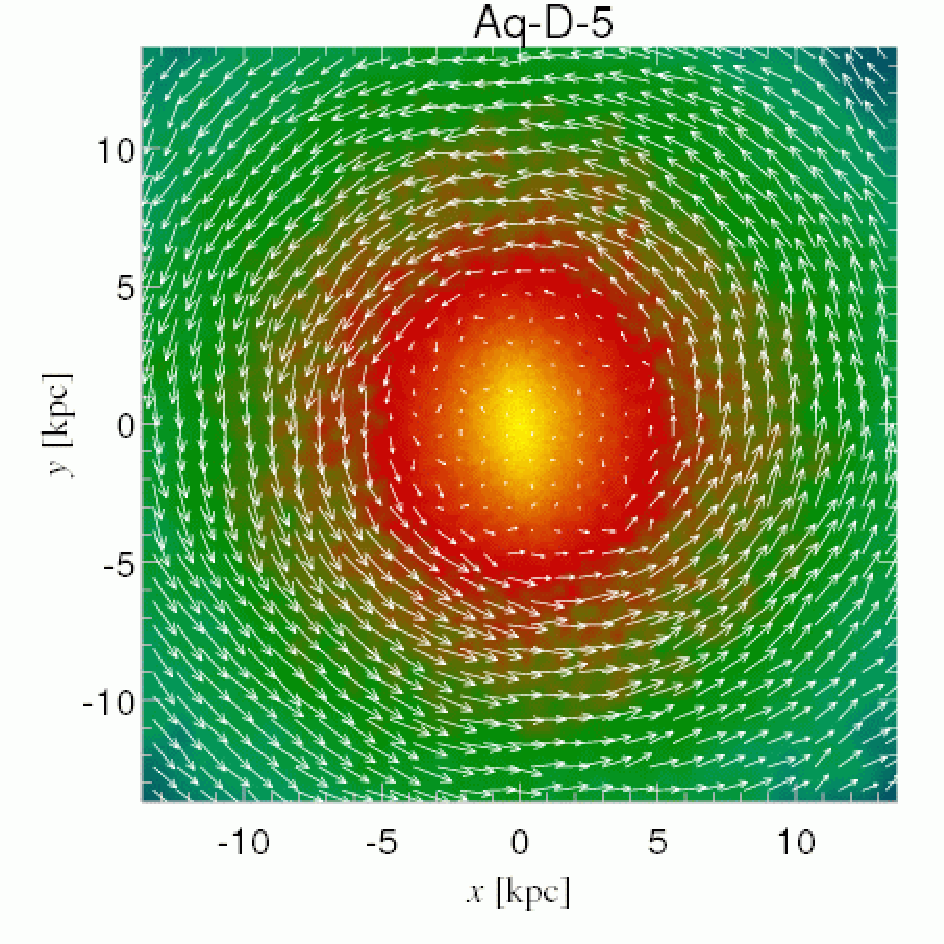}\includegraphics[width=40mm]{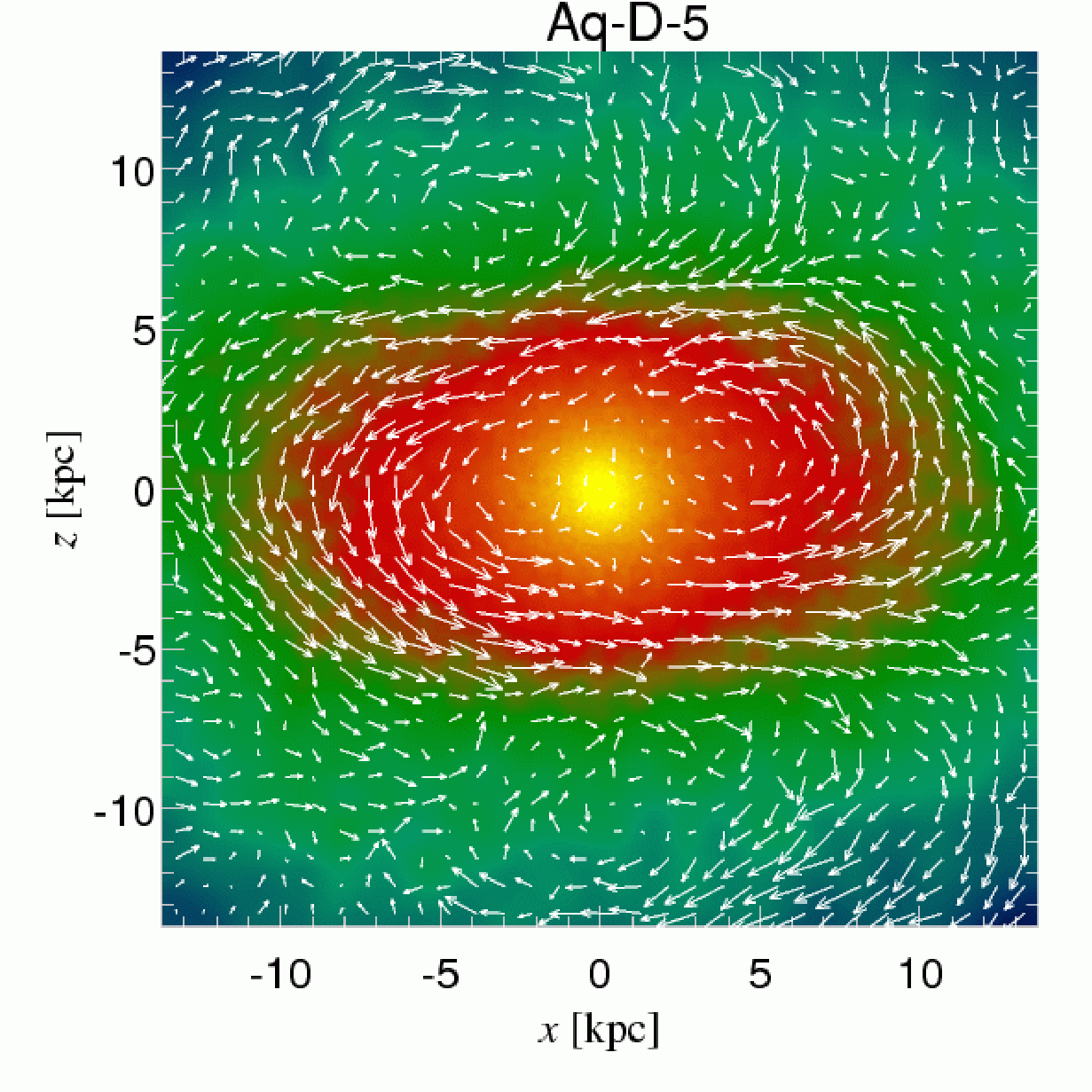}\includegraphics[width=40mm]{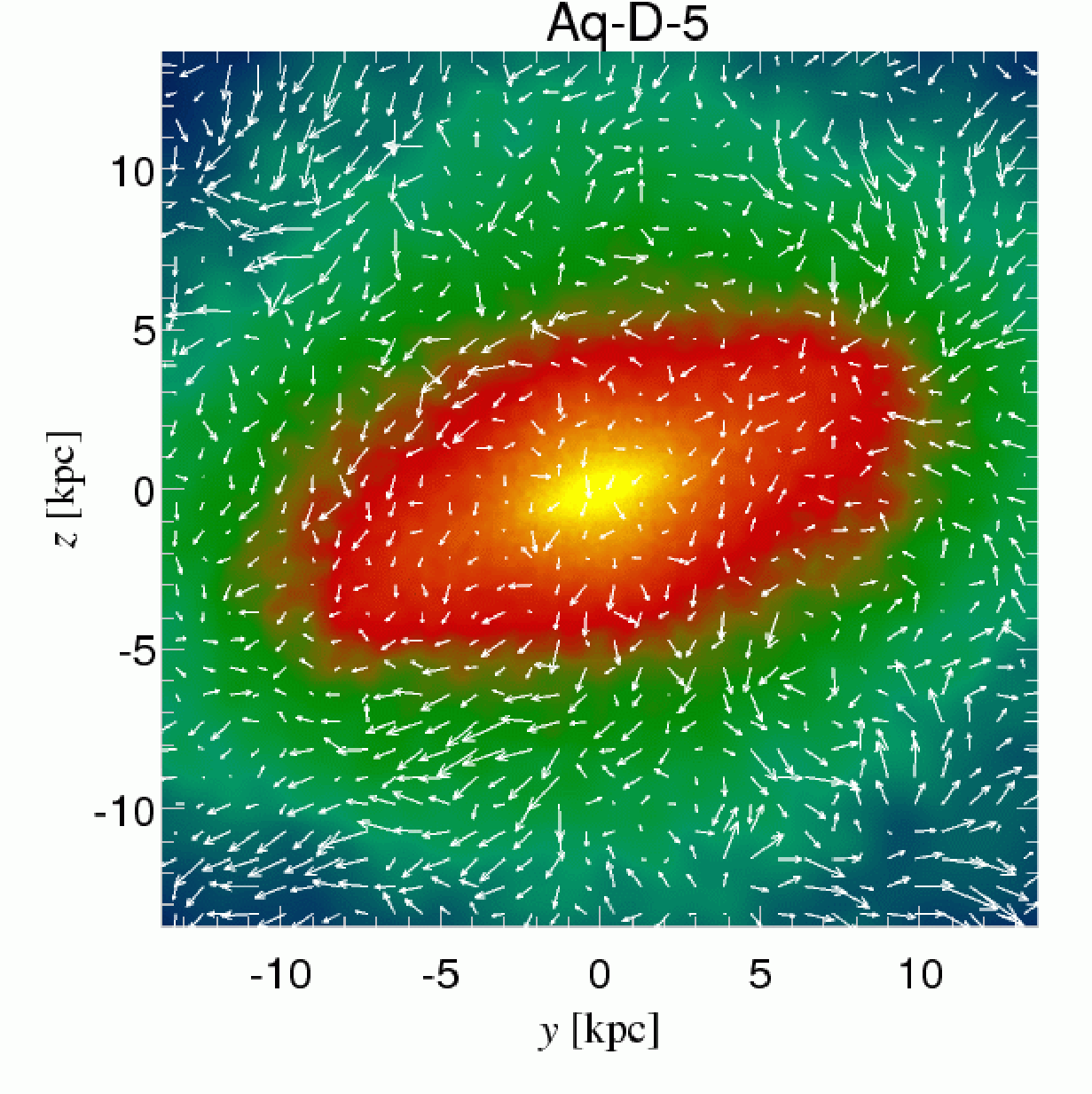}}
\caption{Face-on and edge-on stellar surface density maps for simulations Aq-A-5,
Aq-B-5,  Aq-C-5 and Aq-D-5 at $z=0$.
The colours span $4$ orders of magnitude in projected stellar density, with
brighter colours representing higher densities.
The arrows trace the velocity field of the stars.}
\label{fig_morphologies}
\end{figure*}

In this Section, we study the main properties of the simulated
galaxies. In particular, we investigate their morphologies, 
as well as their star formation histories.
We  also study separately the properties of the discs and spheroids, although
in this paper we focus on the study of the discs,
deferring a more detailed investigation of the
spheroids to another paper.

\begin{figure*}
{\includegraphics[width=40mm]{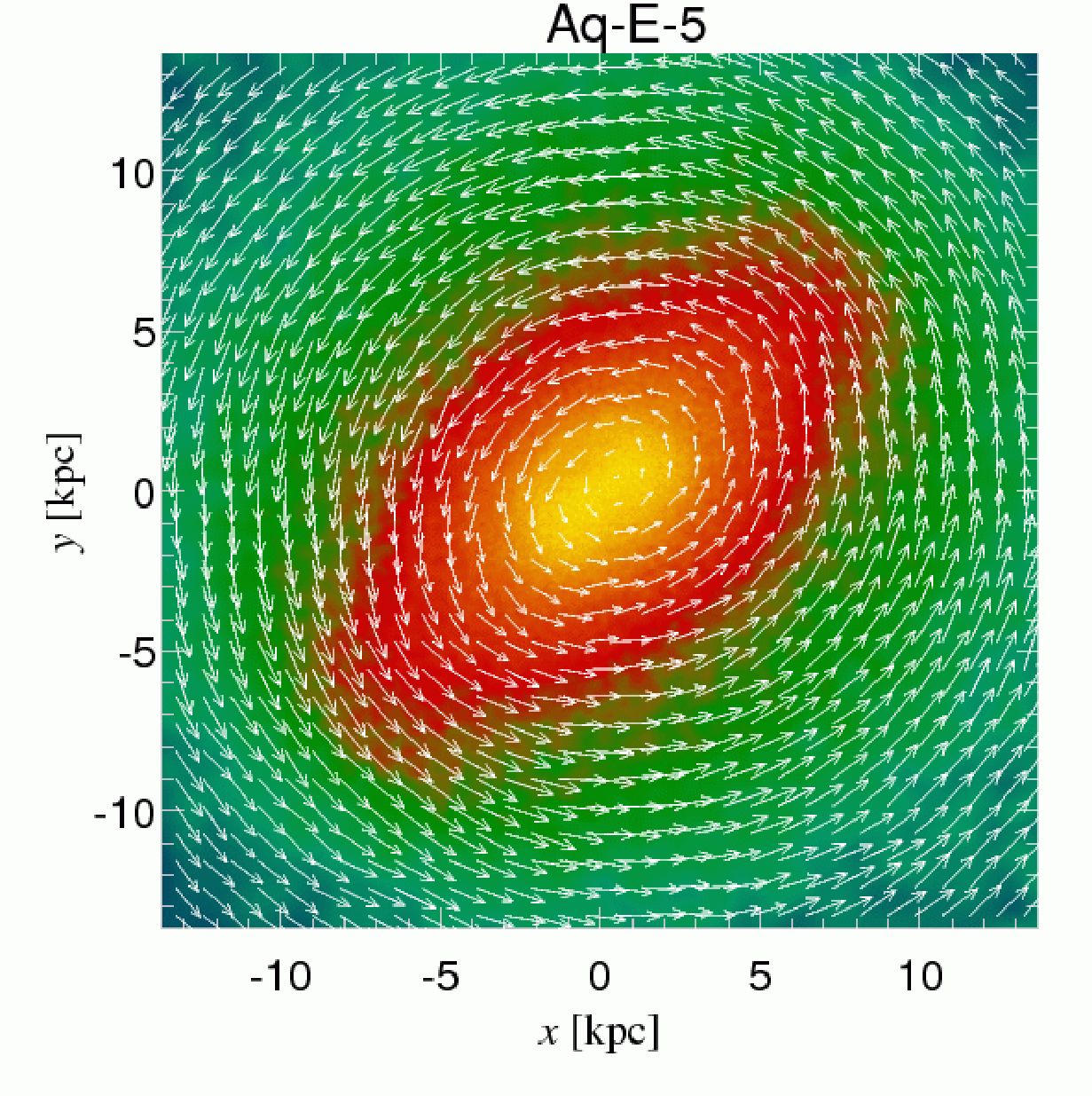}\includegraphics[width=40mm]{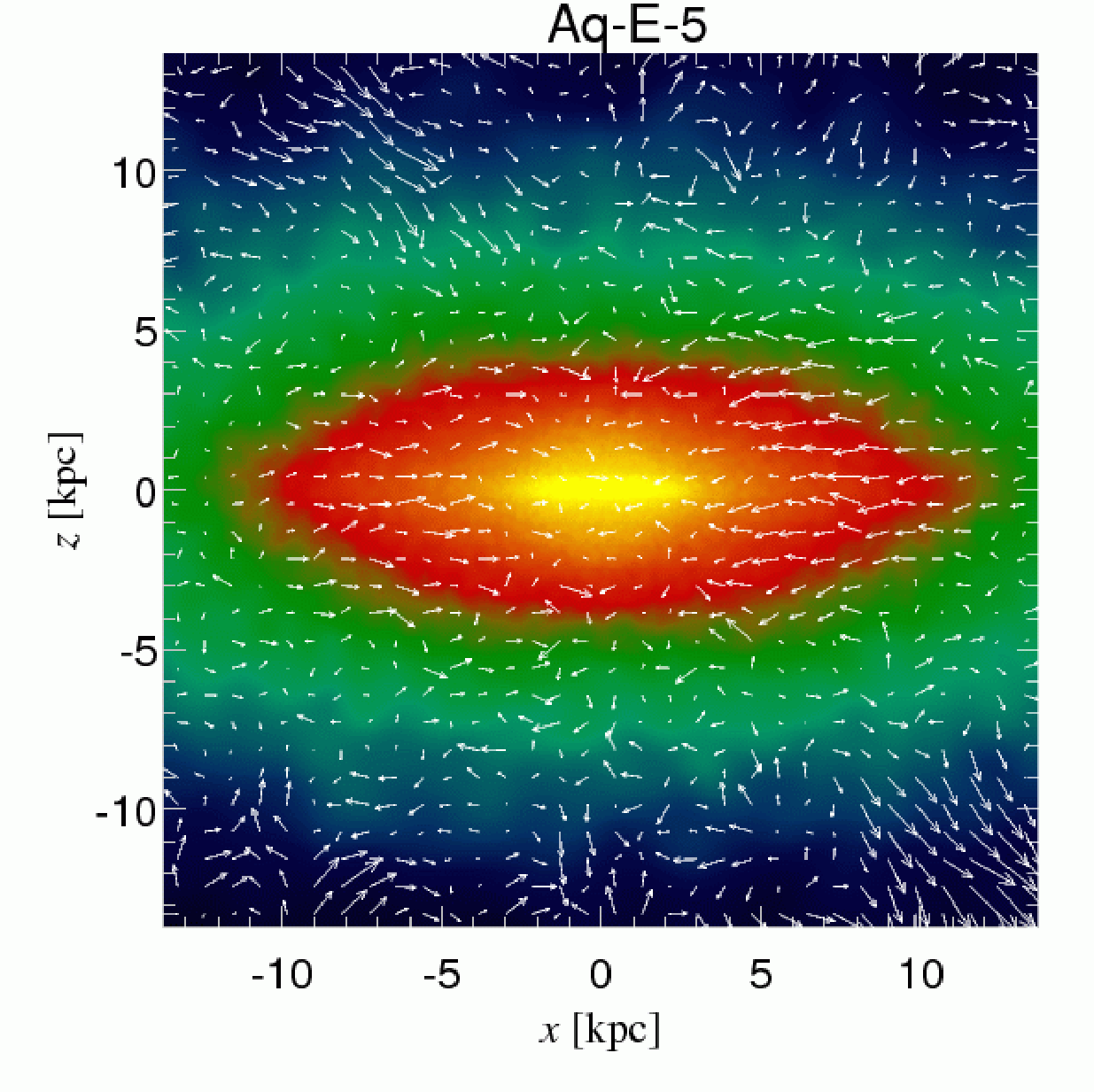}\includegraphics[width=40mm]{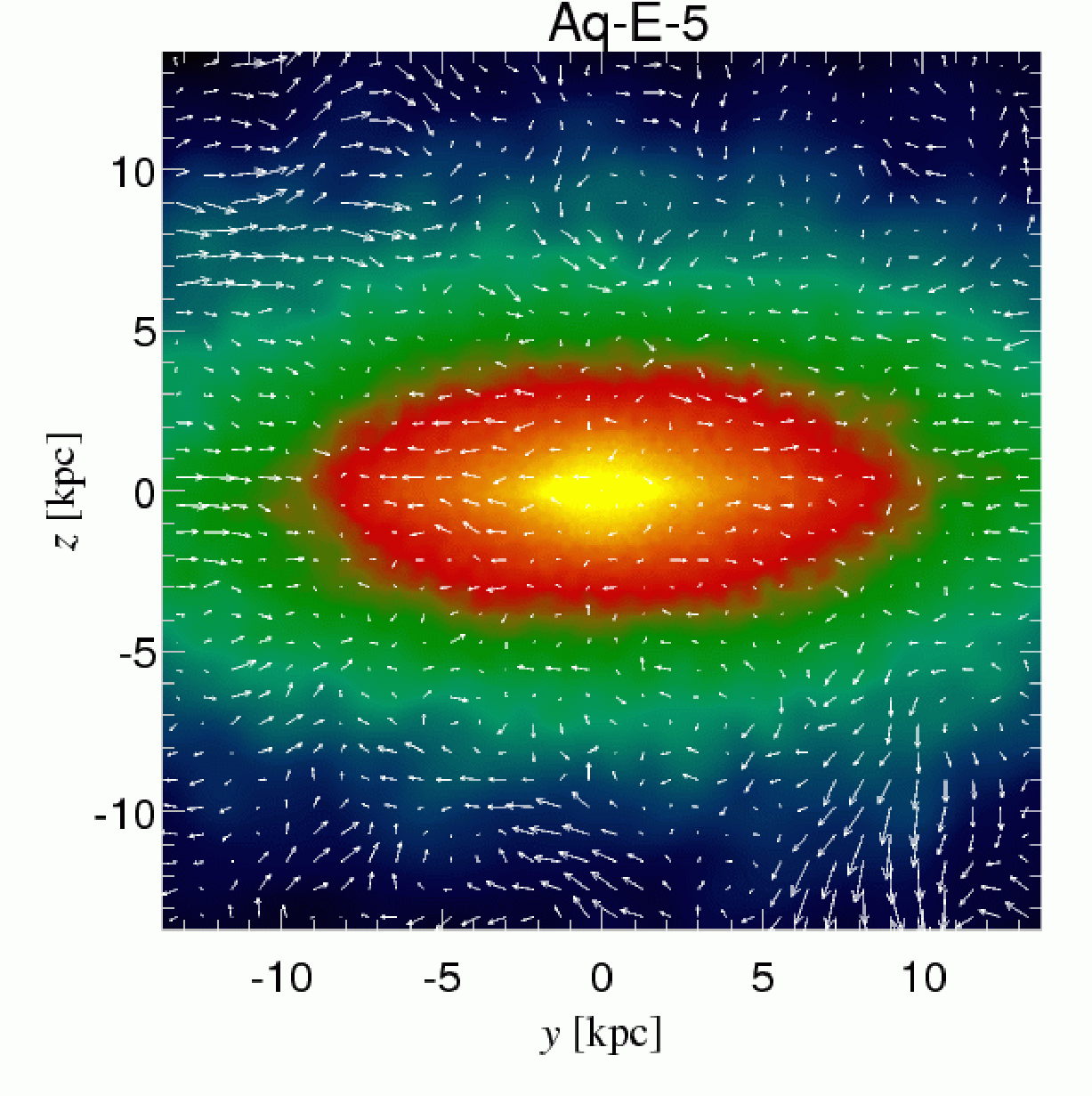}}
{\includegraphics[width=40mm]{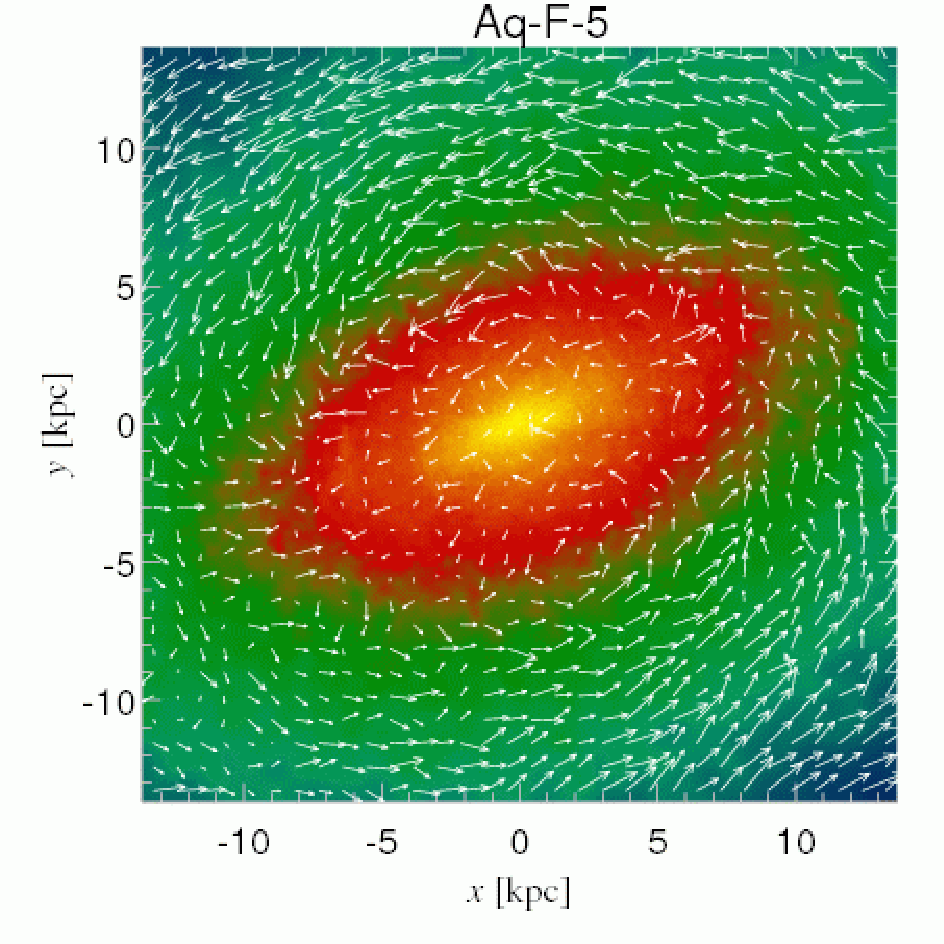}\includegraphics[width=40mm]{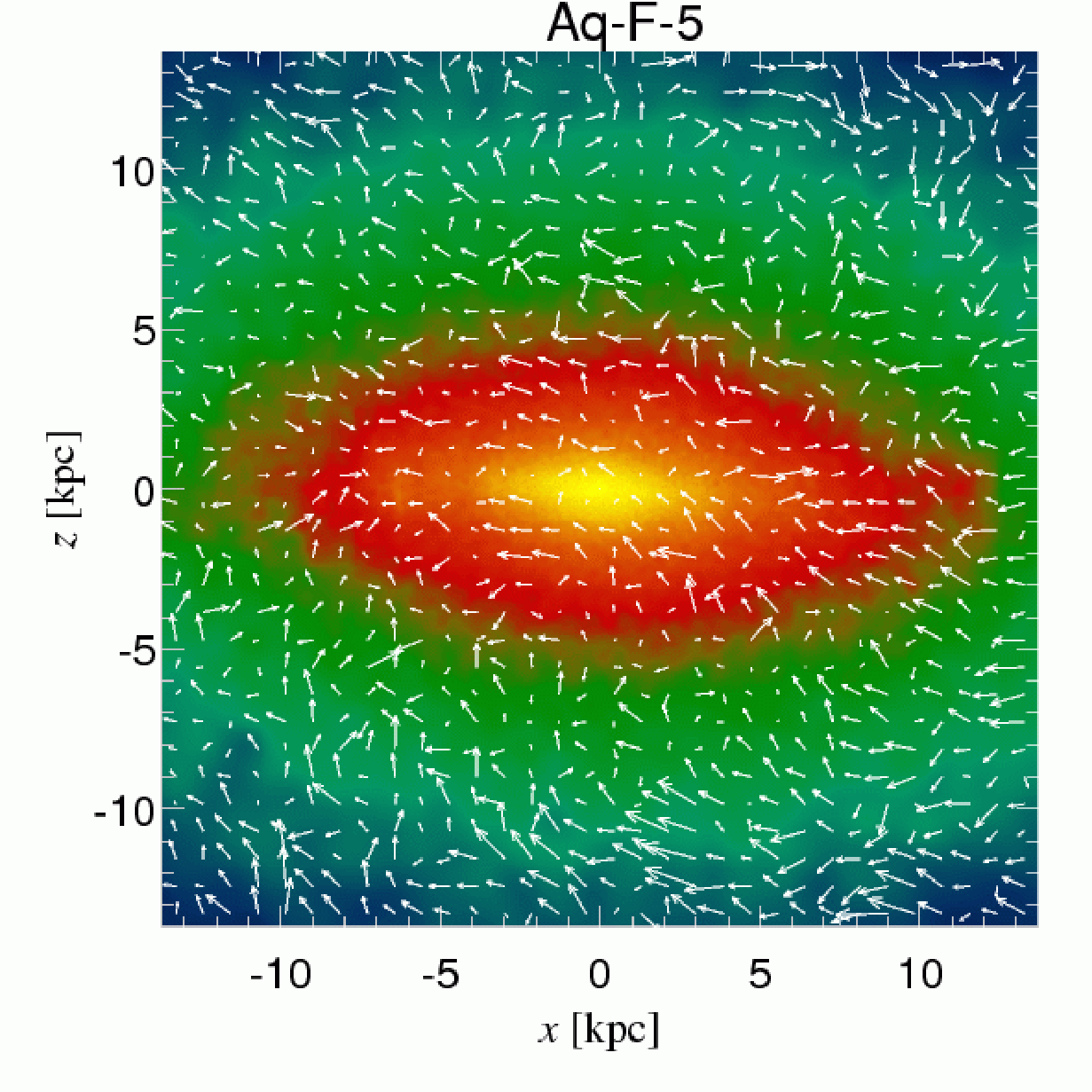}\includegraphics[width=40mm]{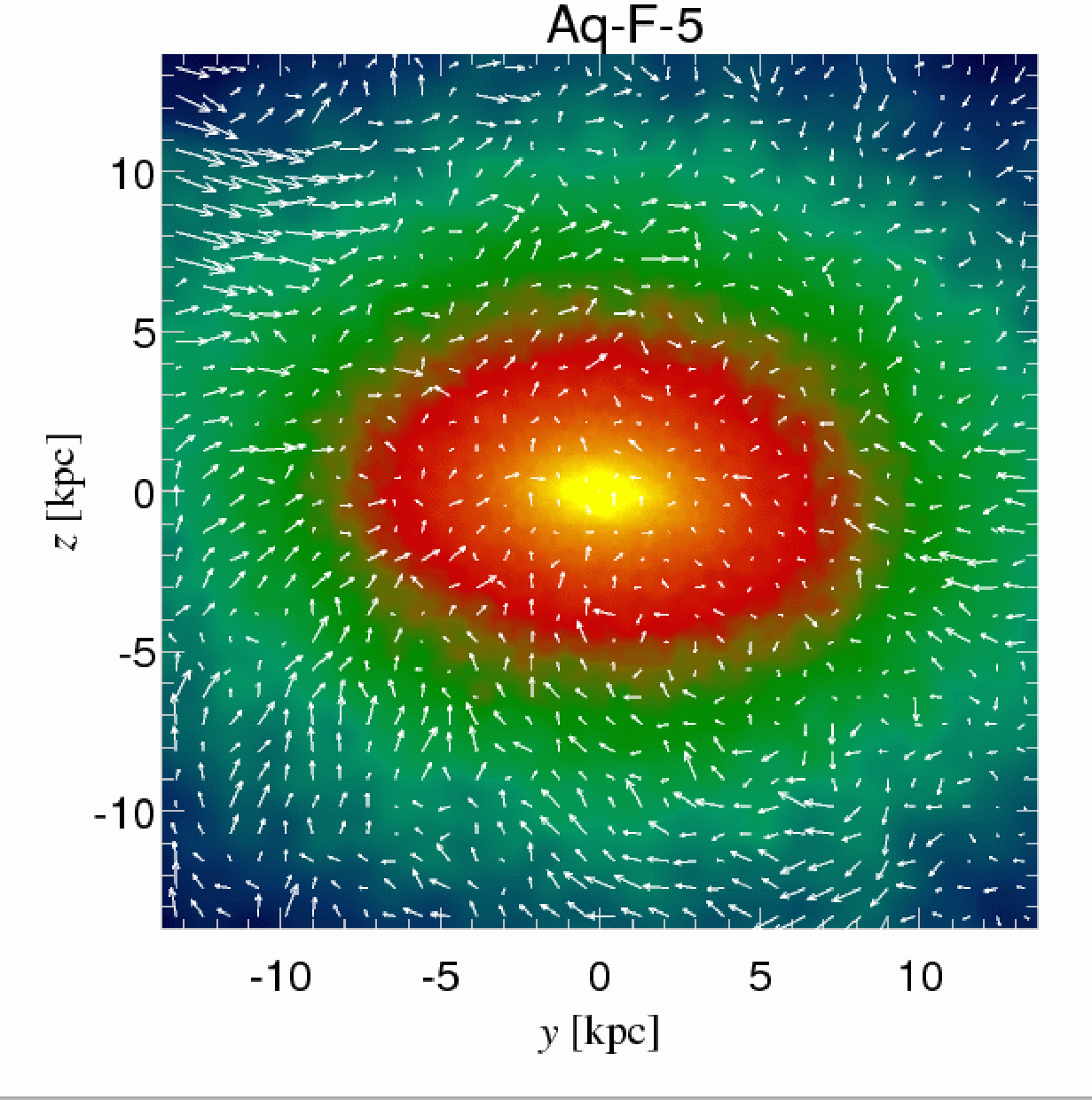}}
{\includegraphics[width=40mm]{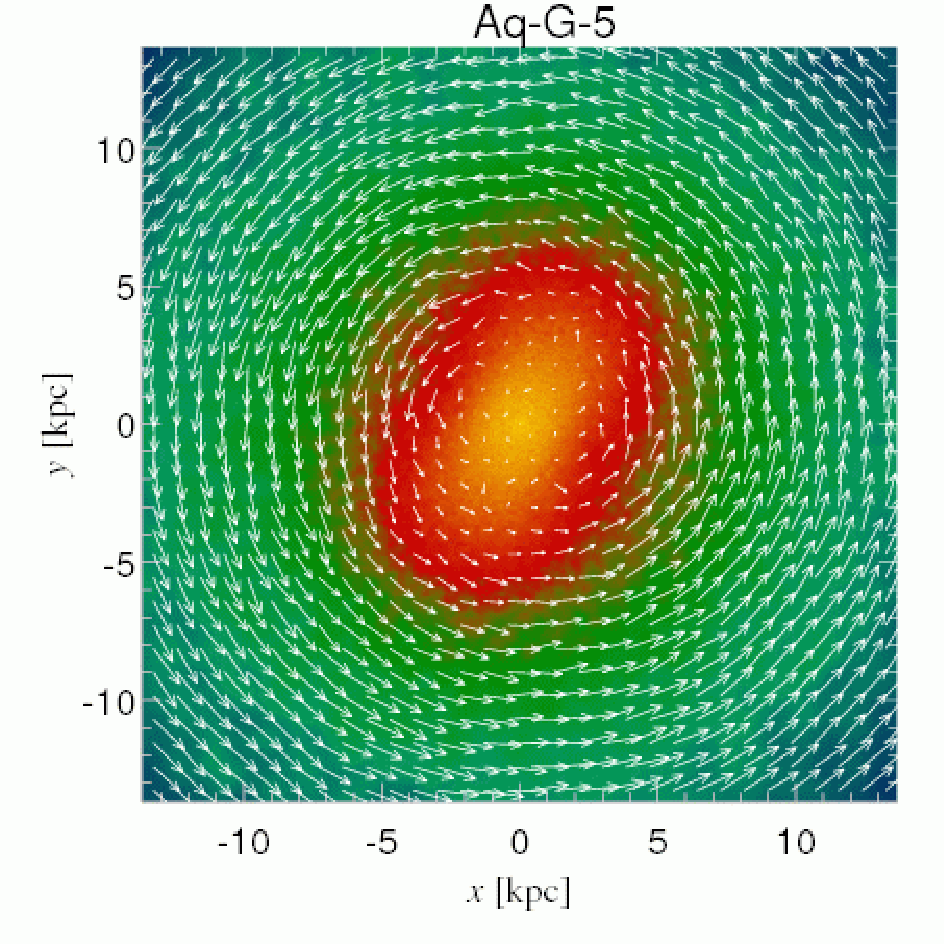}\includegraphics[width=40mm]{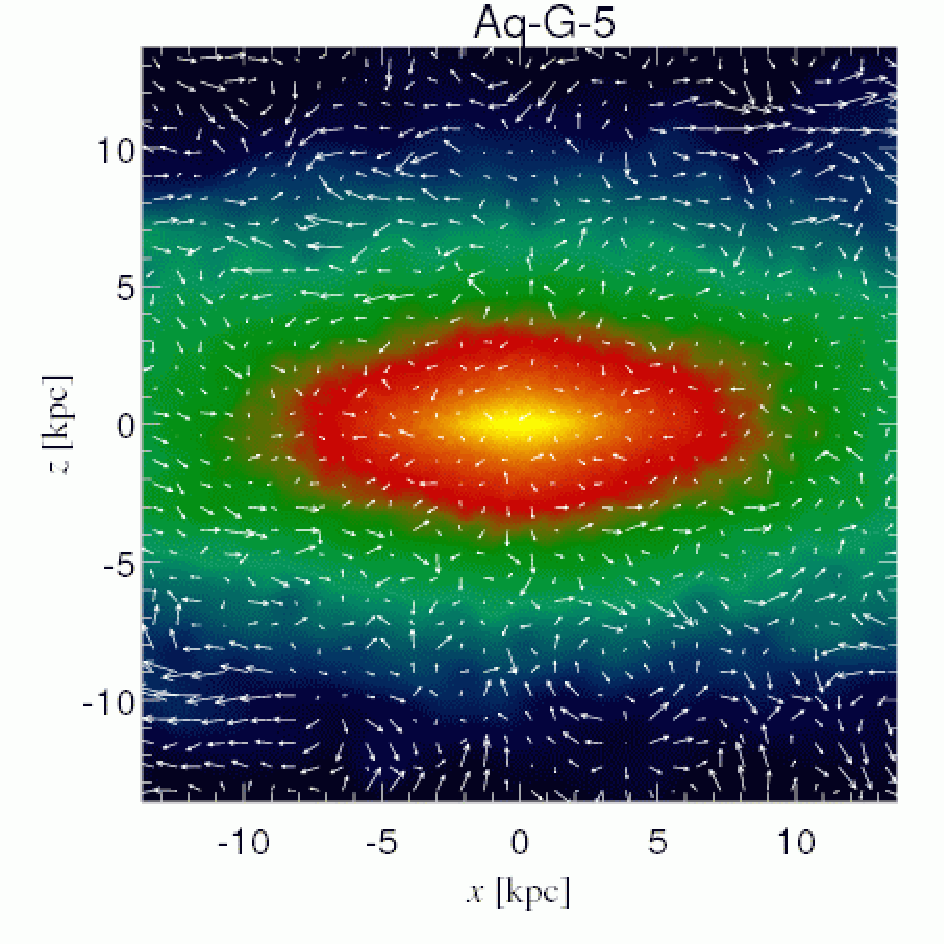}\includegraphics[width=40mm]{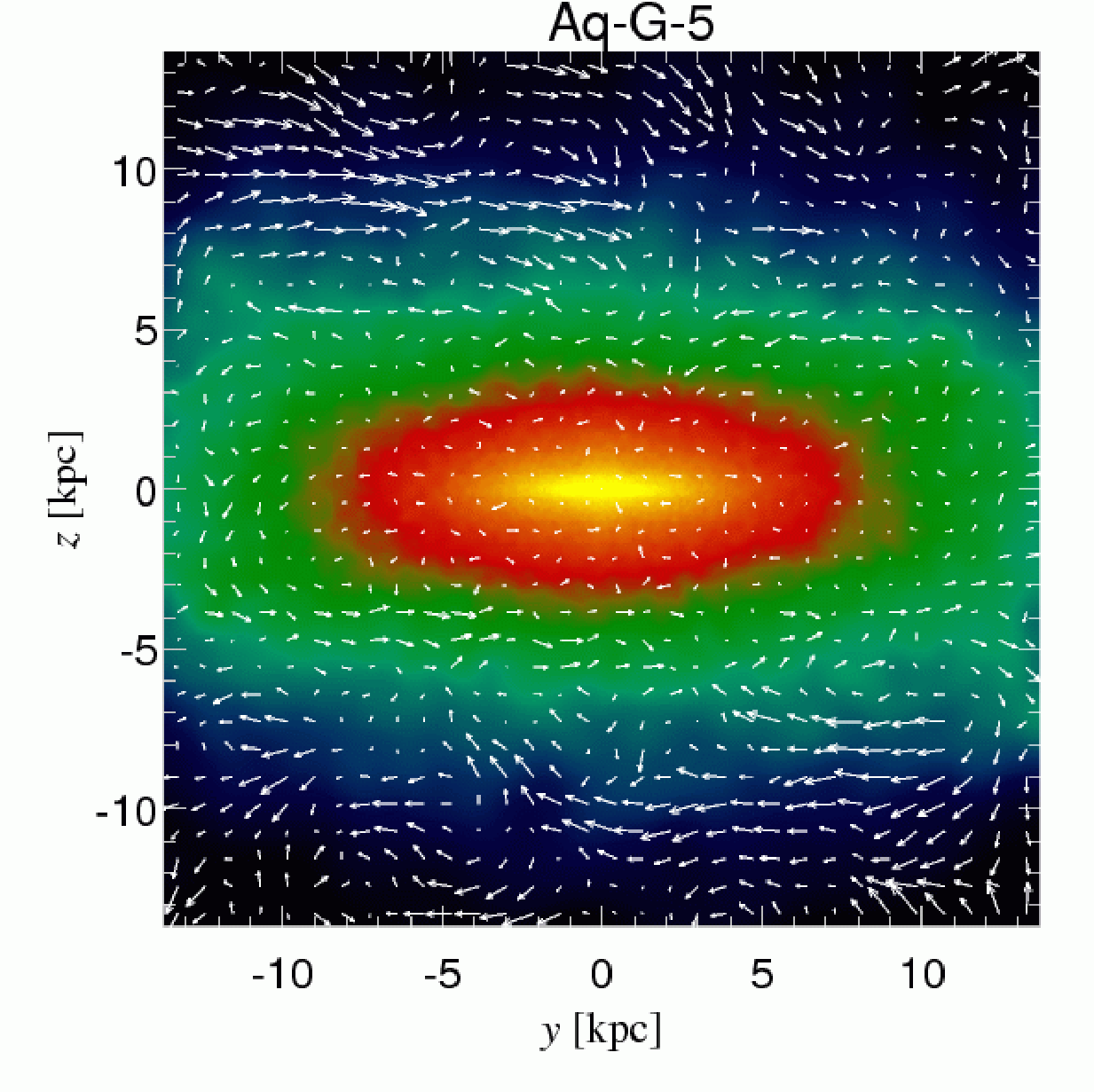}}
{\includegraphics[width=40mm]{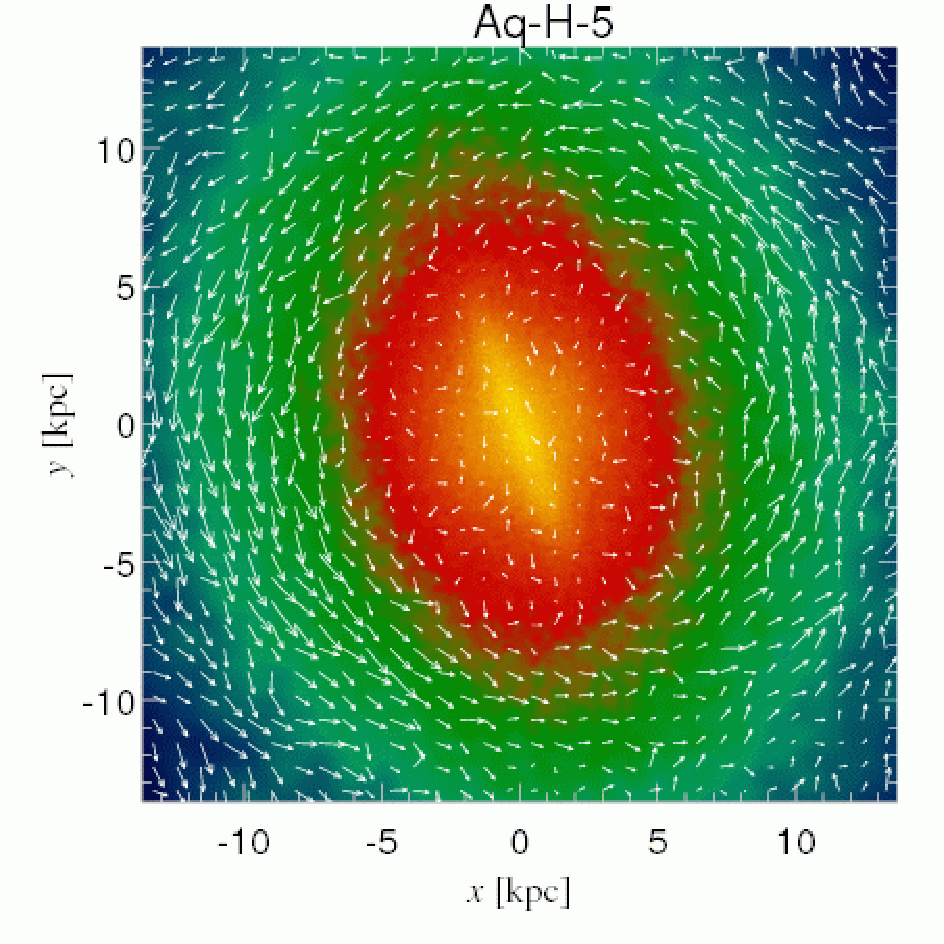}\includegraphics[width=40mm]{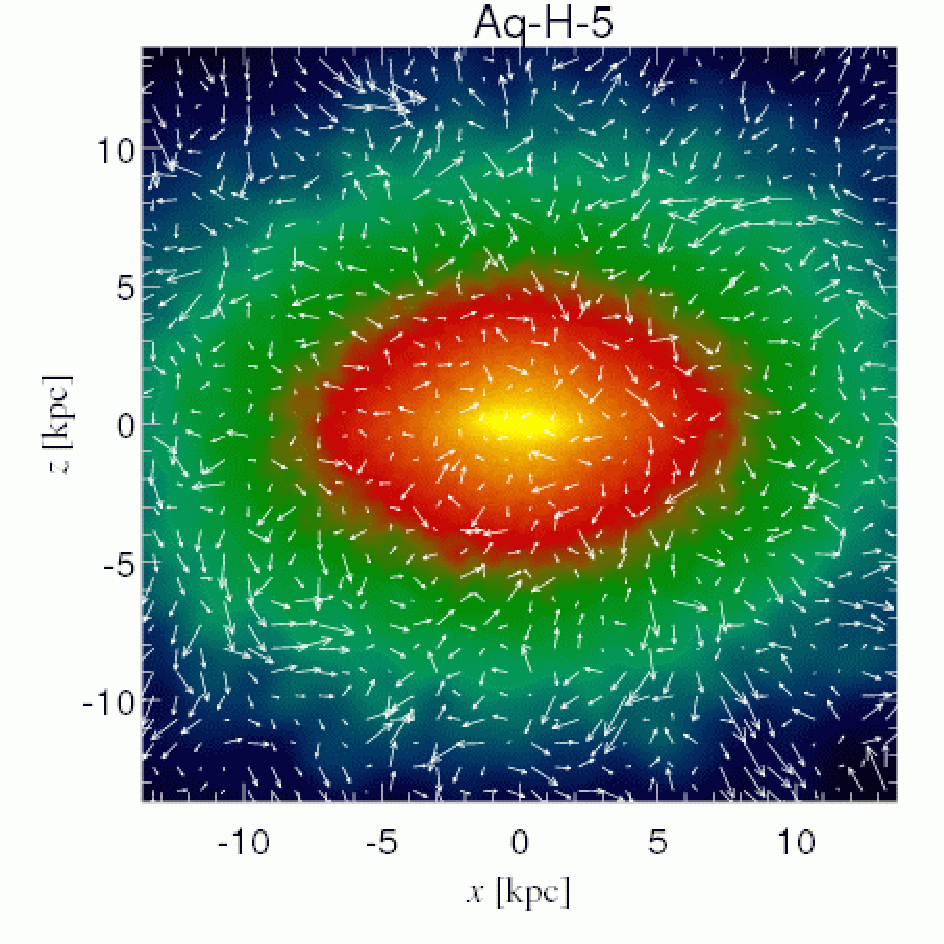}\includegraphics[width=40mm]{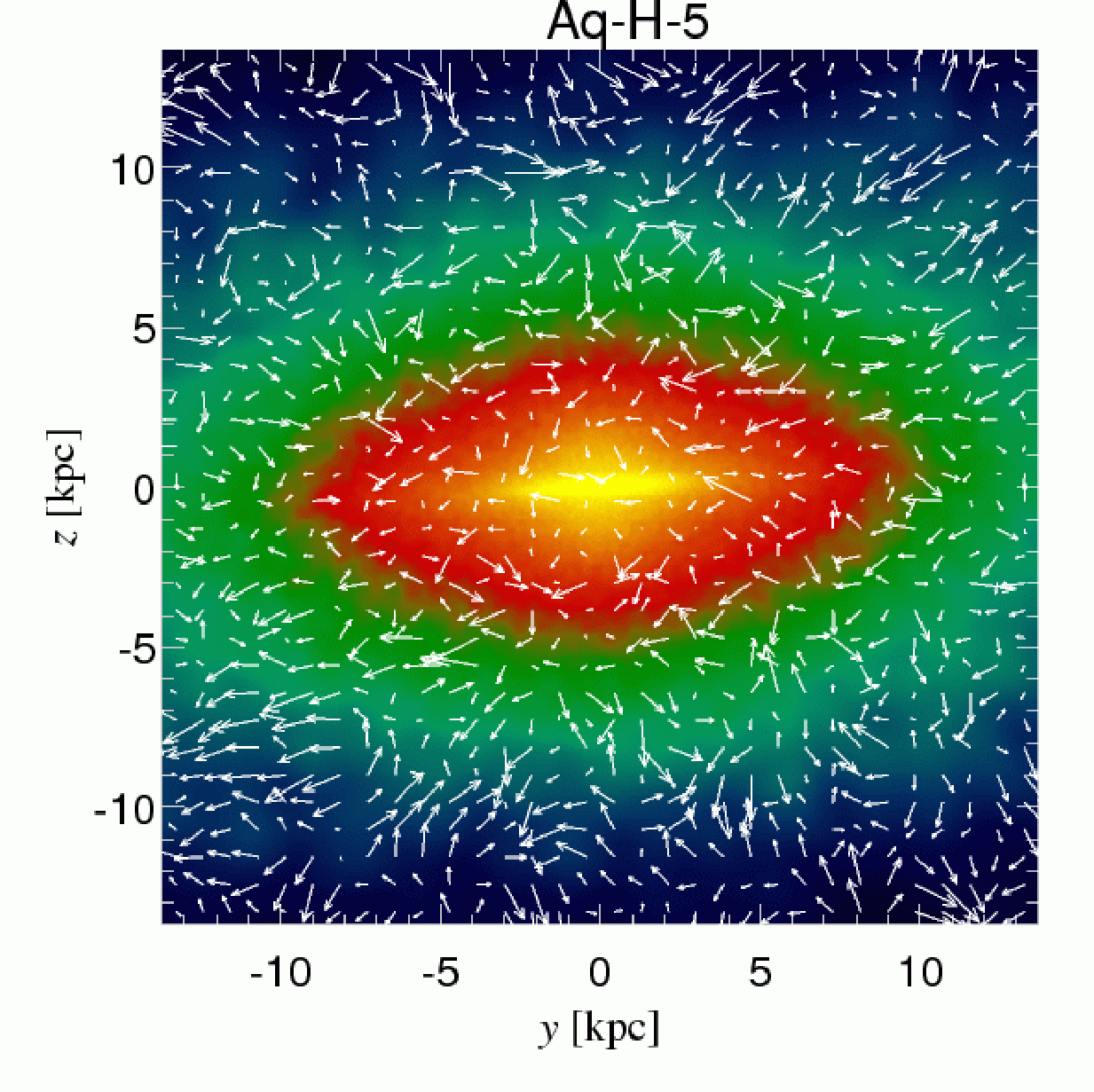}}
\caption{Face-on and edge-on stellar surface density maps for  simulations Aq-E-5, Aq-F-5,
  Aq-G-5 and  Aq-H-5 at $z=0$.
The colours span $4$ orders of magnitude in projected stellar density, with
brighter colours representing higher densities.
The arrows trace the velocity field of the stars.}
\label{fig_morphologies2}
\end{figure*}

In order to study the morphologies of the simulated
galaxies  at $z=0$, we calculate  the angular
momentum of baryons in the inner regions of the haloes  $\vec{J}$,
and rotate  the reference system 
in order to align  $\vec{J}$  with the $z-$axis.
Figs.~\ref{fig_morphologies} and ~\ref{fig_morphologies2}  show maps
of the stellar densities projected onto planes perpendicular and parallel to 
the $z$ direction.
Left-hand panels
represent  face-on views, whereas the  middle and right-hand panels
correspond to  edge-on projections.
The colours span $4$ orders of magnitude in projected density,
with brighter colours representing higher densities.
Arrows trace the corresponding projected velocity fields.
We find that the eight simulated galaxies present 
a variety of morphologies. 
It is striking that none of them has a dominant thin disc.
About half of them do have substantial disc components,
reflected both in the projected mass  distributions
and in their velocity fields, which display
ordered rotation. 
An interesting feature is that, except for Aq-E-5, 
discs do not seem to extend to the innermost regions
(see also below).
Some of the simulated galaxies 
show little or no disc component, and no
clear sign of ordered rotation around any axis.
Also, we find that bars are present in most cases;
the simulated galaxies are never perfectly axisymmetric.

\begin{figure*}
{\includegraphics[width=42mm]{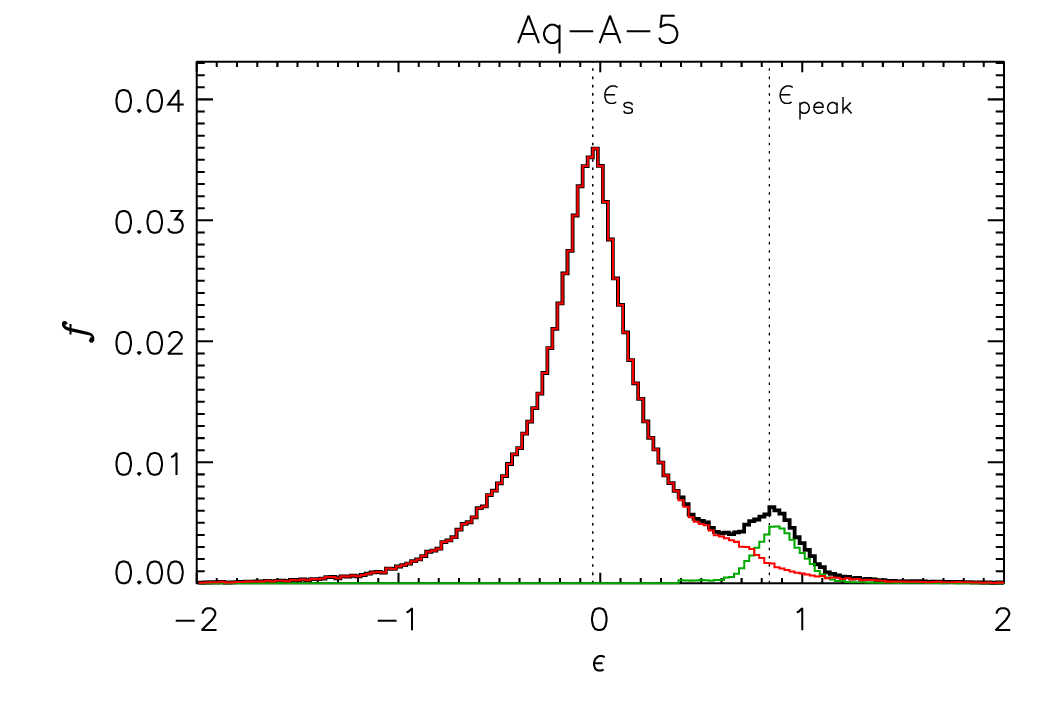}\includegraphics[width=42mm]{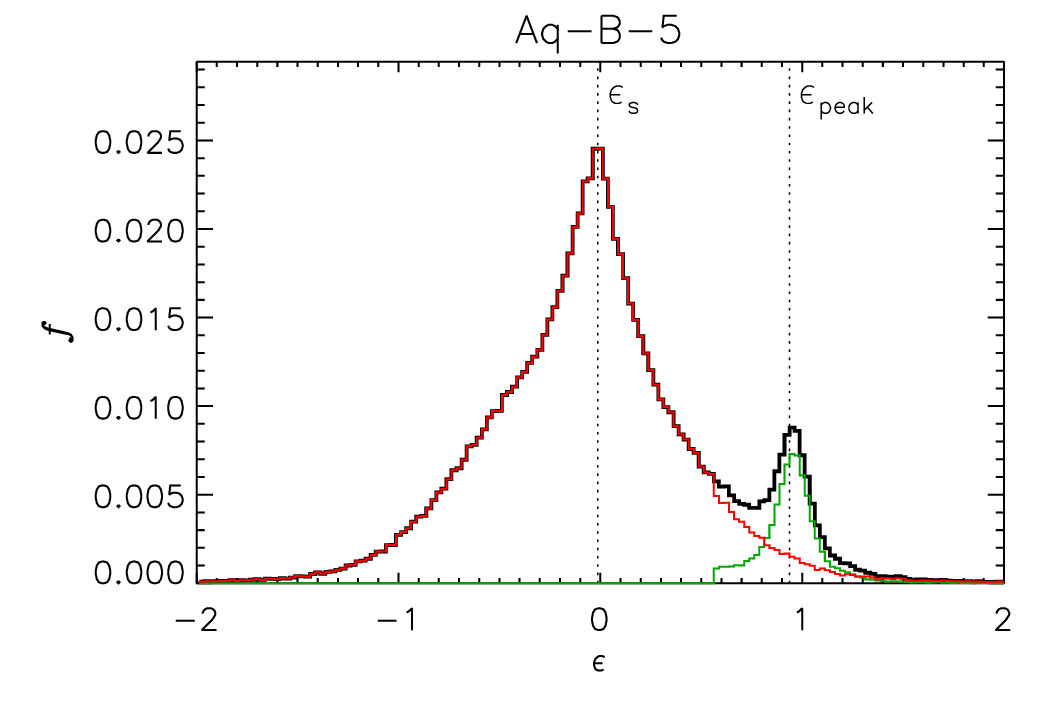}\includegraphics[width=42mm]{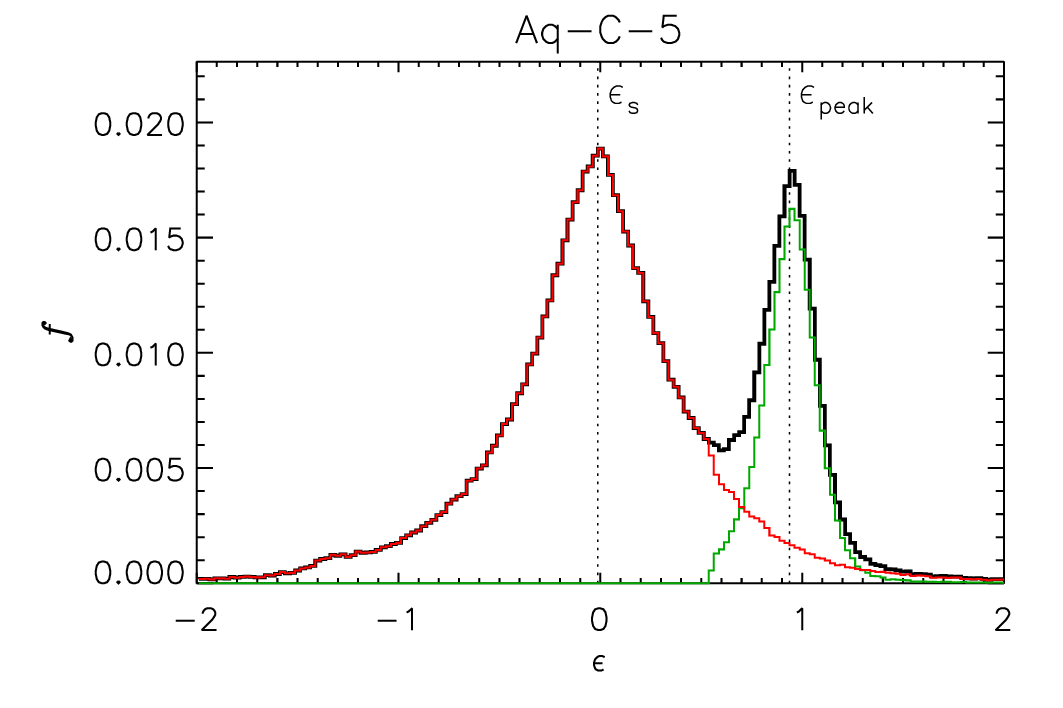}\includegraphics[width=42mm]{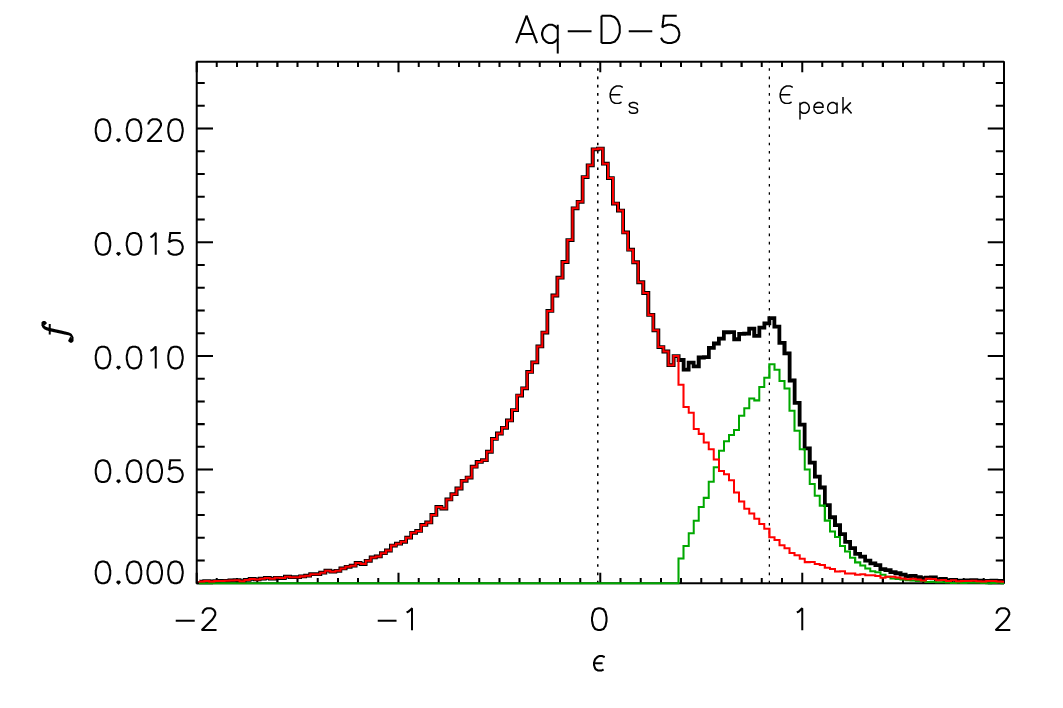}}
{\includegraphics[width=42mm]{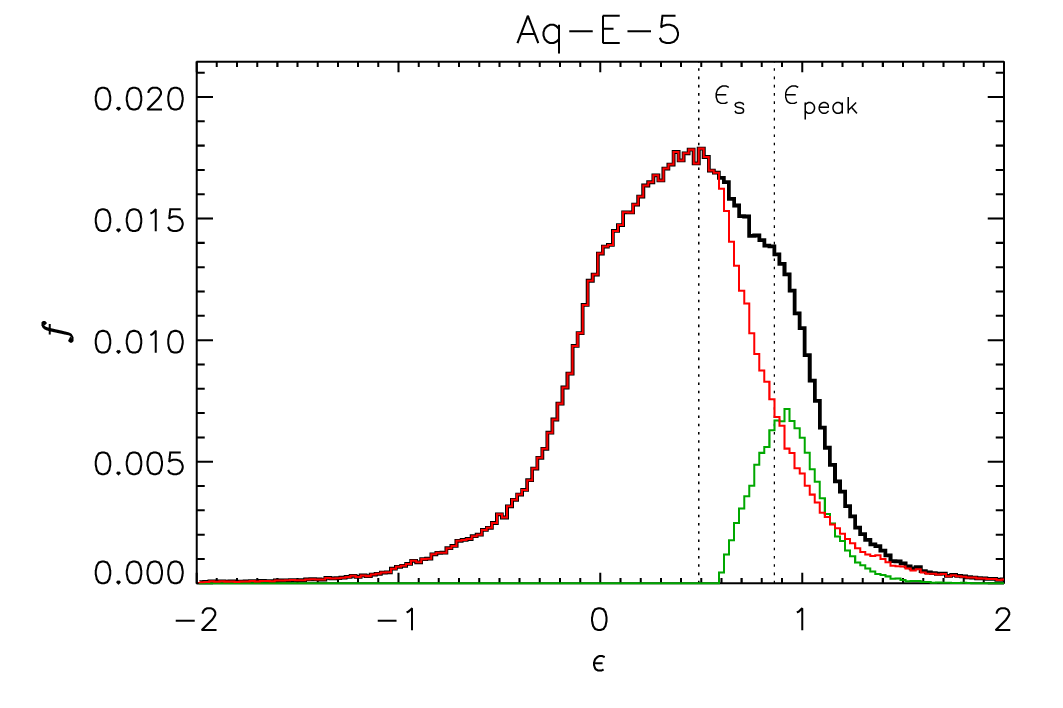}\includegraphics[width=42mm]{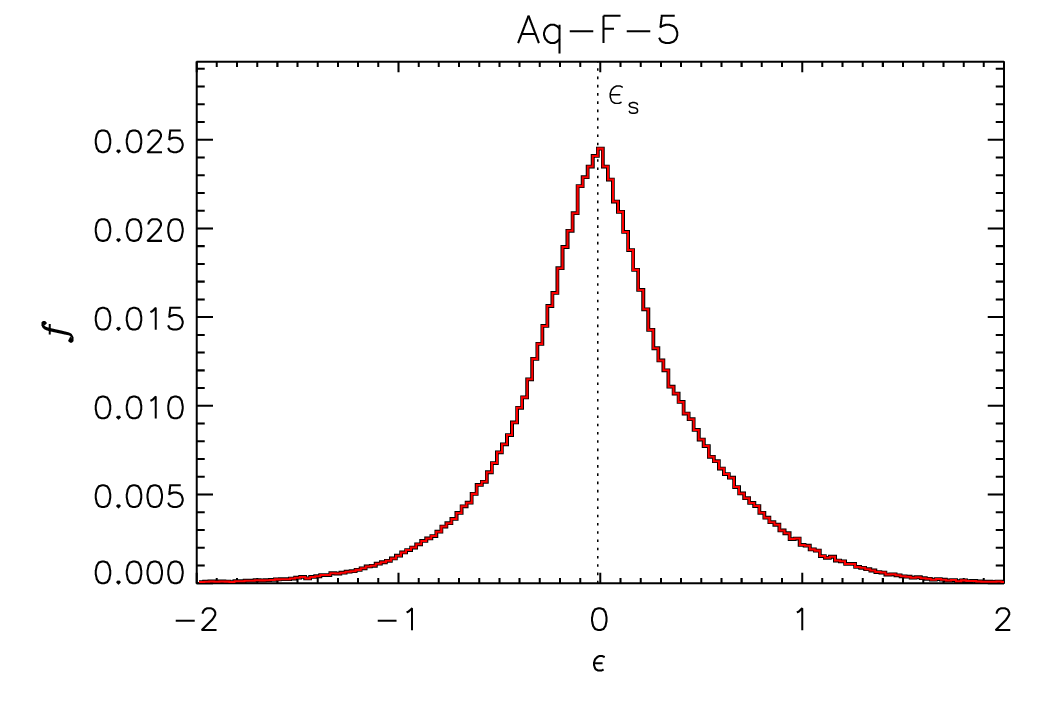}\includegraphics[width=42mm]{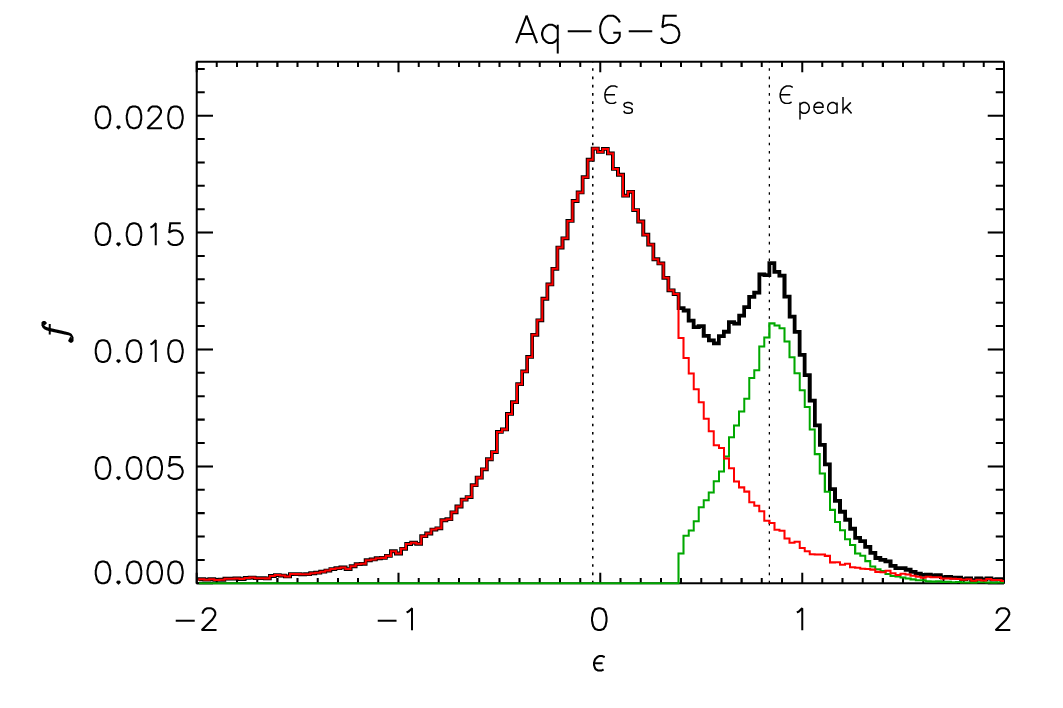}\includegraphics[width=42mm]{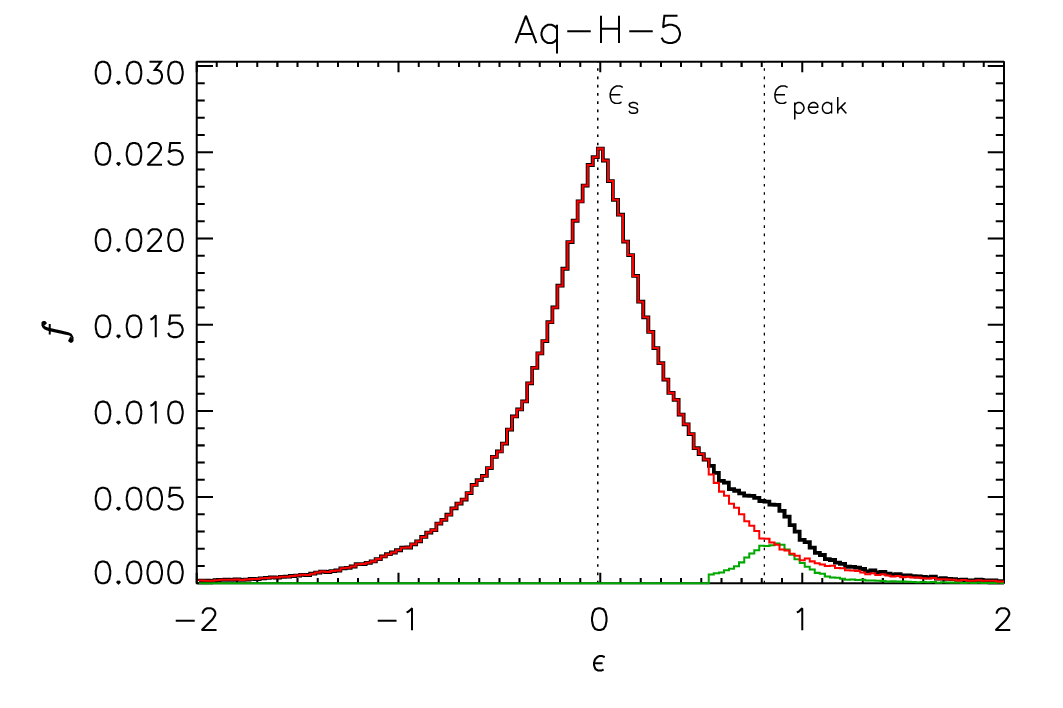}}
\caption{Stellar mass fraction as a function of $\epsilon=j_z/j_{\rm circ}$
for our set of simulations  at $z=0$ (black lines). 
Red and green lines correspond to the distributions for spheroid and disc stars, respectively,
obtained with our disc-spheroid decomposition. 
We only consider stars which belong to the main system
(i.e. we discount stars in satellites).}
\label{fig_epsilon}
\end{figure*}

\begin{figure*}
{\includegraphics[width=42mm]{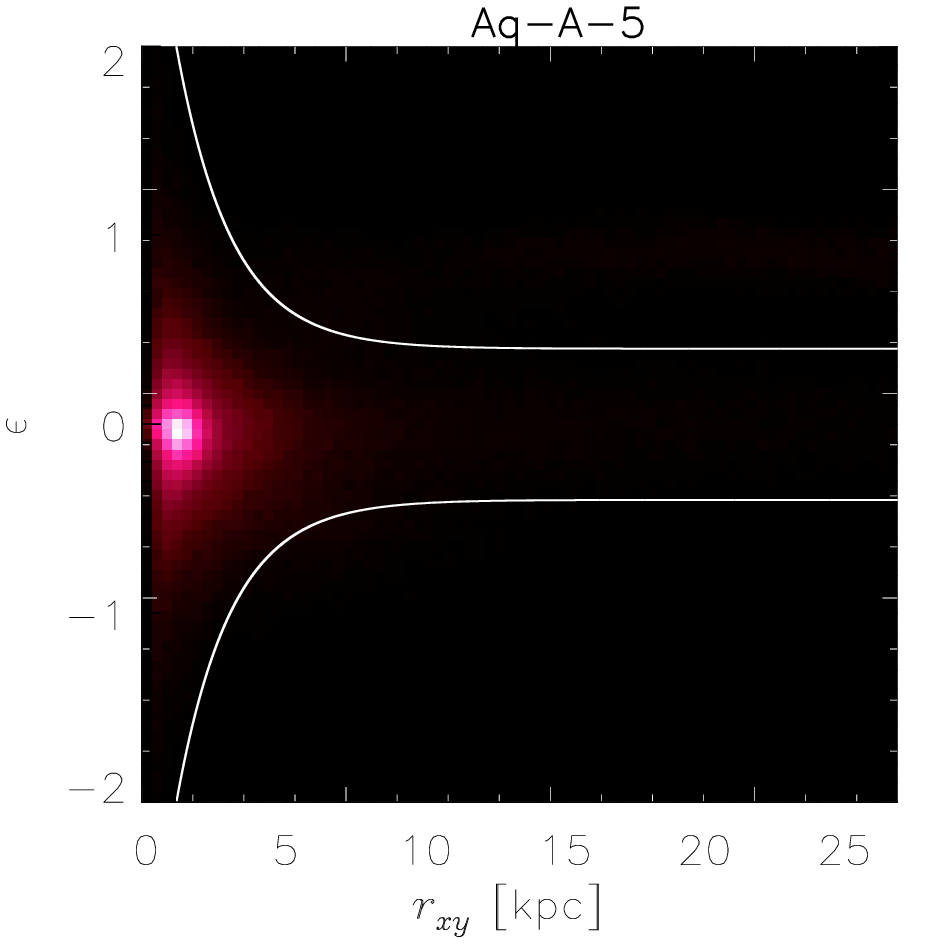}\includegraphics[width=42mm]{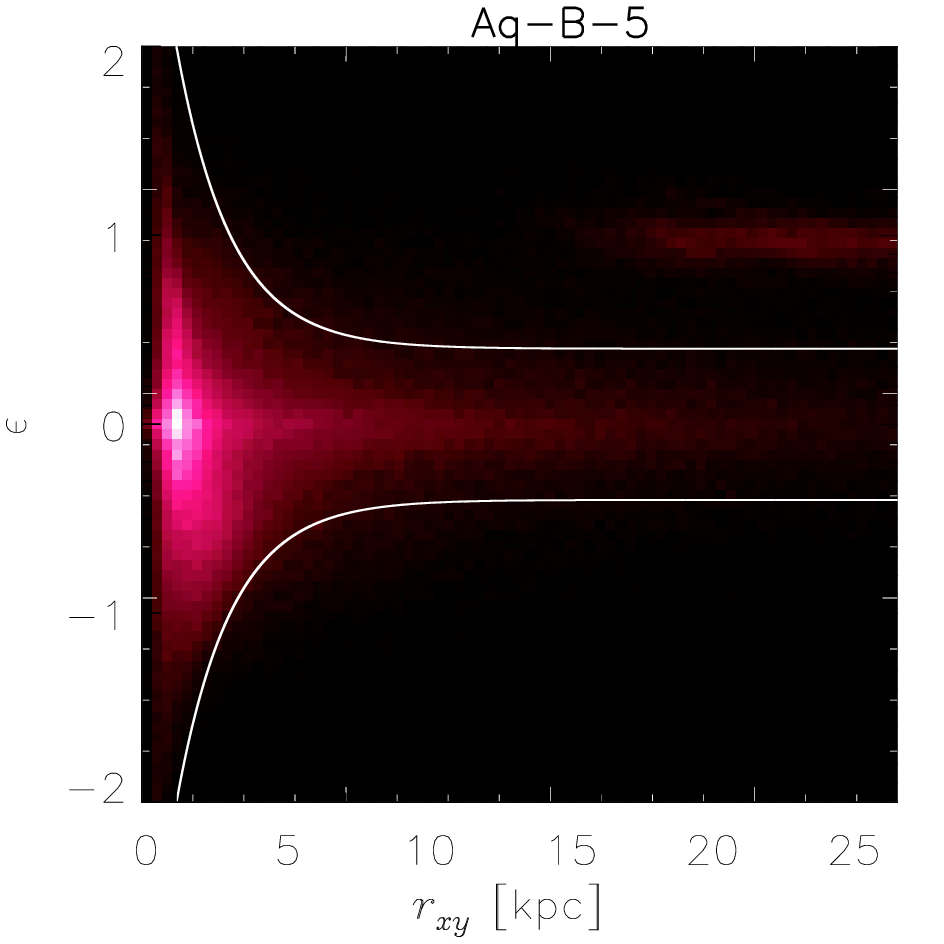}\includegraphics[width=42mm]{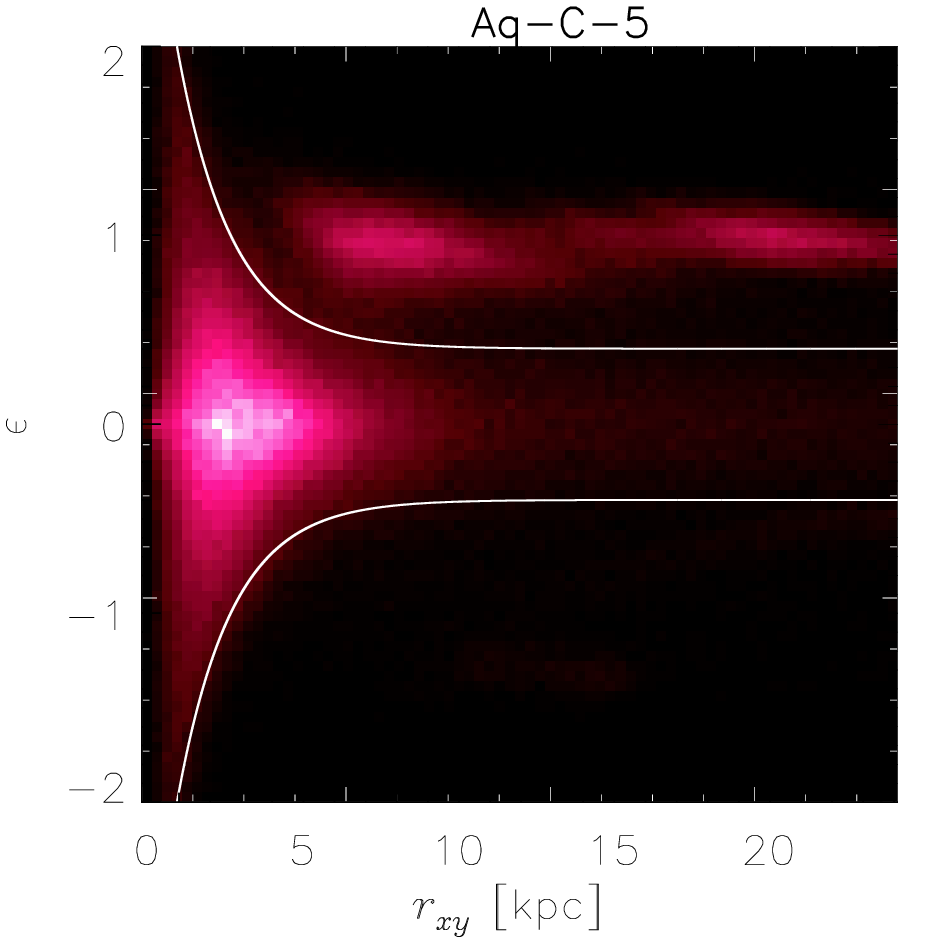}\includegraphics[width=42mm]{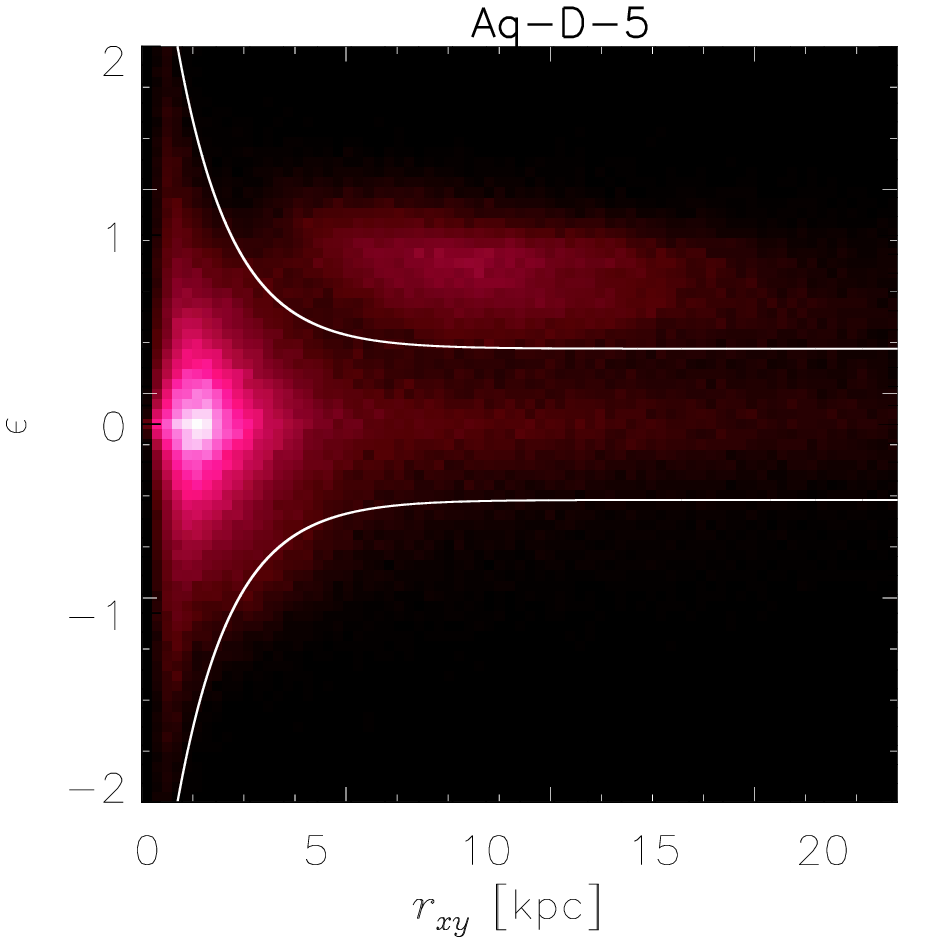}}
{\includegraphics[width=42mm]{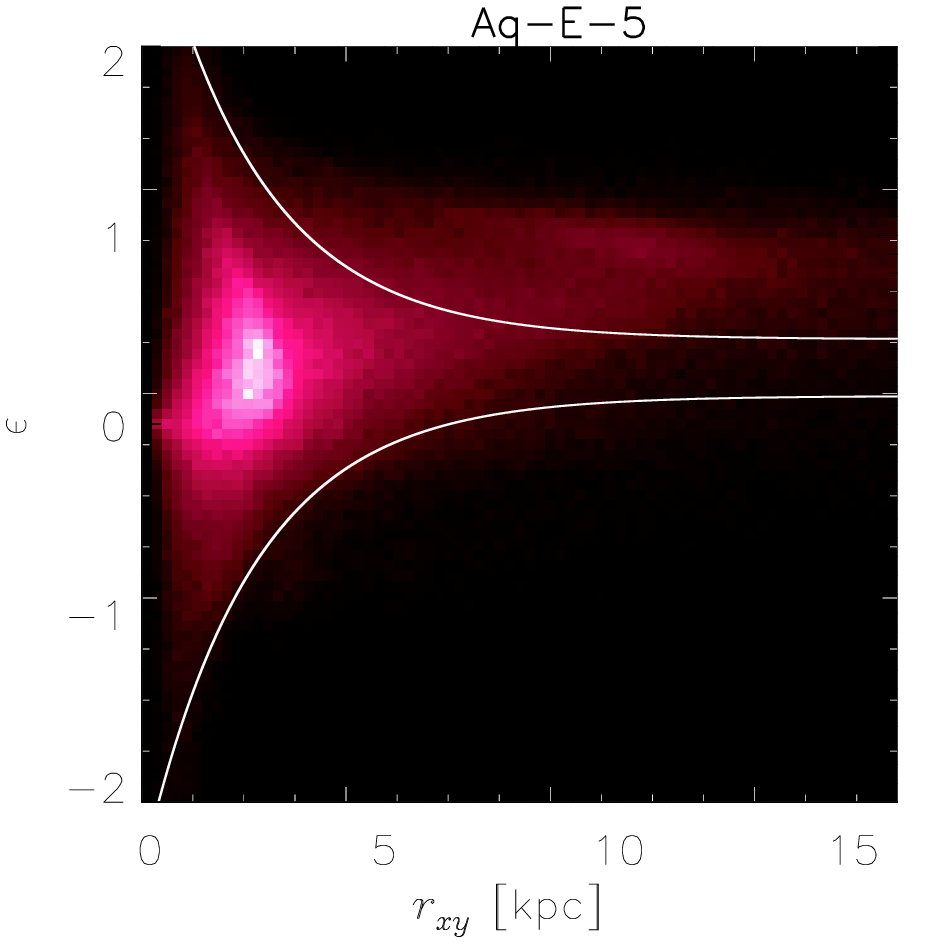}\includegraphics[width=42mm]{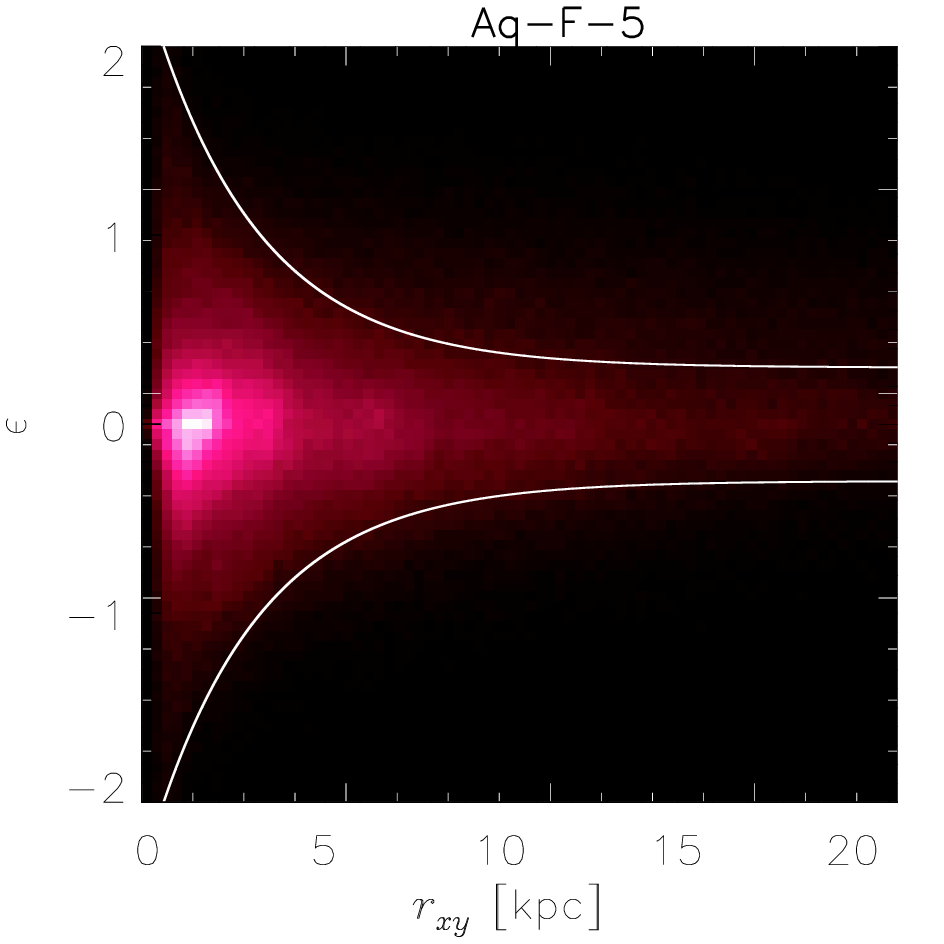}\includegraphics[width=42mm]{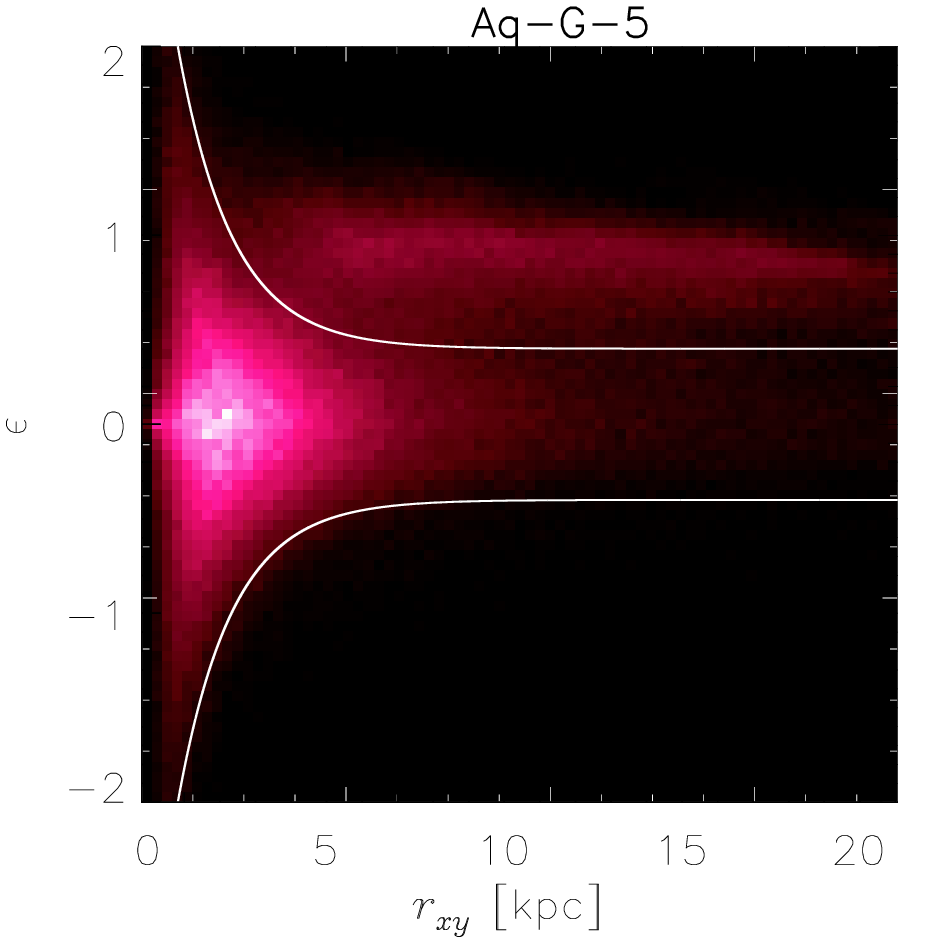}\includegraphics[width=42mm]{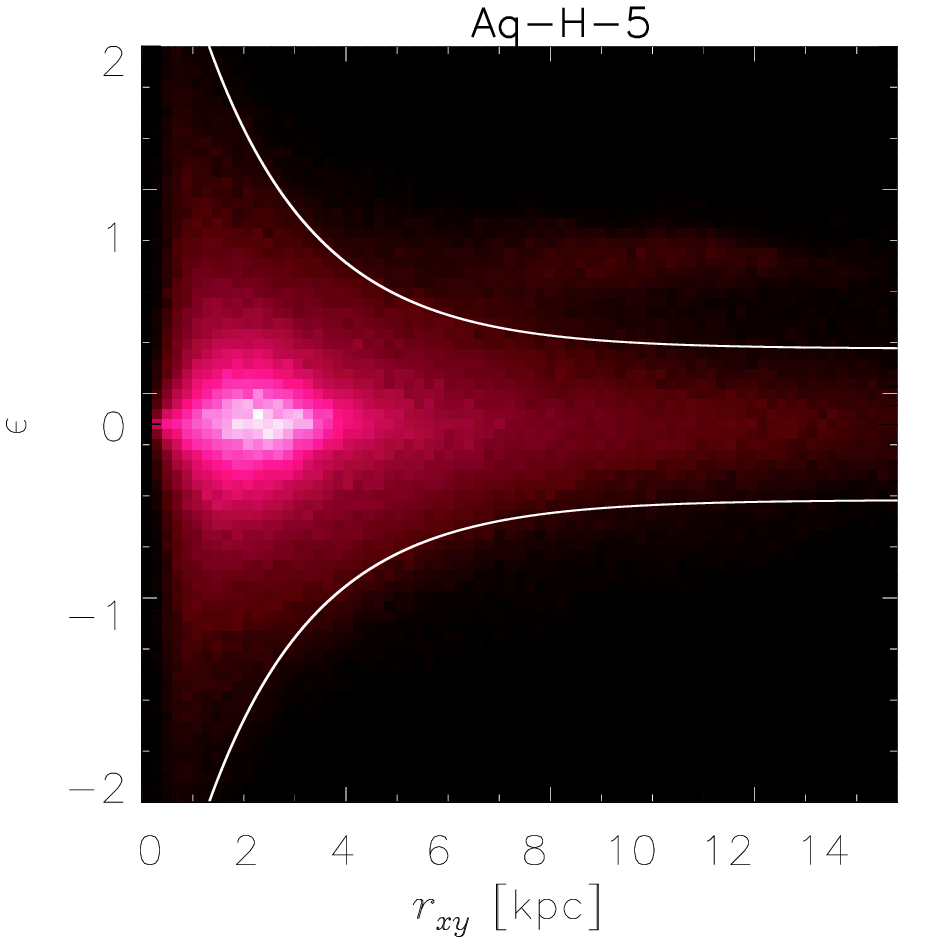}}
\caption{Eccentricity, $\epsilon=j_z/j_{\rm circ}$, as a function of radius in the disc plane
for our set of simulations at $z=0$.
The colours scale with the density of points in each pixel, and brighter
colours correspond to higher densities.}
\label{fig_epsilon_vs_r}
\end{figure*}

In order to characterize the properties of the discs and spheroids
formed
in our simulations, we designed a procedure to segregate
stars into two main components,
which we refer to as {\it disc} and {\it spheroid}.
This is similar to the method used in S08, 
but is improved in order to avoid contamination
of the disc by spheroid stars, particularly in the central regions.
Our procedure has a series of steps. First, we
construct the distributions of the parameter
$\epsilon$ of the simulated stars,
defined as $\epsilon=j_z/j_{\rm circ}$, where  $j_z$
is the angular momentum of each star in the $z$-direction
(i.e. the direction of $\vec{J}$), and
$j_{\rm circ}$ is the angular momentum expected for a circular orbit  at the same
radius: $j_{\rm circ}= r \cdot v_{\rm circ}(r)$ (we use $v_{\rm circ}(r) 
= \sqrt{G\ M(r) / r}$, although the mass distribution is not
spherically symmetric in the central regions).
The parameter $\epsilon$  
can be used to  distinguish kinematically between
stars in the disc and in the spheroid.
In Fig.~\ref{fig_epsilon}, we show 
its distribution for our  simulated galaxies.
We note that for this analysis, we only consider particles which belong
to the main  system  within
twice the optical radius,  ignoring stars in satellites.
The distributions typically show two  peaks:
one at $\epsilon_{\rm peak}\sim 1$ ($\epsilon_{\rm peak}$  changes slightly between
 galaxies)
which is indicative of the presence of a disc in rotational support, 
and a second
one at  $\epsilon\sim 0$, corresponding to  a spheroidal
distribution dominated by velocity dispersion.
We define $\epsilon_{\rm s}$ as the location of this peak, as indicated in Fig.~\ref{fig_epsilon_vs_r}.
Note that $\epsilon_{\rm s}\sim 0$ for all simulations except Aq-E-5,
where $\epsilon_{\rm s} \sim 0.45$, indicating that this
spheroid has a substantial net rotation.

From the distribution of $\epsilon$ for the simulated galaxies,
we can infer the relative importance of discs and spheroids.
From our eight galaxies, four of them clearly have well-defined
disc components (Aq-C-5, Aq-D-5, Aq-E-5 and Aq-G-5), 
three of them have dominant spheroids and small
discs (Aq-A-5, Aq-B-5 and Aq-H-5), and 
one is a spheroidal system with no disc at all (Aq-F-5).
Clearly, the $\epsilon$ distributions are useful to assign
stars either to the disc or the spheroidal components.
However, we find that a definition
based on $\epsilon$ alone can lead to 
contamination of the discs by spheroid stars.
We therefore looked for additional constraints that allow
a better segregation of  discs from spheroids.
In  Fig.~\ref{fig_epsilon_vs_r},  we show the distribution
of $\epsilon$ as a function of projected distance in the disc plane
(up to $1.5\times r_{\rm opt}$) for our simulations. 
The colours represent the density of particles in each pixel.
Clearly, there are typically two distinct components which populate
different regions of this plane: disc stars 
($\epsilon\sim 1$) are more extended
in radius but do not seem to extend into the innermost regions, 
while  spheroids are symmetrically located
around $\epsilon\sim 0$  (except
for Aq-E-5) and their stars are more concentrated within the
central regions.
We use this feature to help decide if
stars belong to the disc component,
by requiring disc stars to be located above
the upper white lines shown in Fig.~\ref{fig_epsilon_vs_r}\footnote{ 
See Appendix~\ref{ap_A} for details in the disc-spheroid decomposition 
and in the procedure to construct the white lines.
}.
We also require disc stars to have orbits 
roughly contained in the disc plane. This is done by restricting disc stars
to have cos $\alpha$$>0.7$,  where $\alpha$ is the angle between the angular momentum
of each star and  $\vec{J}$.
Stars which do not fulfil the conditions explained
above are considered part of the spheroidal components.
Note that the central regions of our galaxies are dominated
by velocity dispersion (Fig.~\ref{fig_epsilon_vs_r}), so stars are present
on all possible orbits.
Our procedure avoids
contamination of the disc by high $\epsilon$ stars
in the central region, which more likely belong
to the spheroidal component. 
Note that we define spheroid stars as those which are not
 part of the rotationally supported disc. As a result, spheroids will typically contain
not only bulge and stellar halo stars, but also
stellar bars, if they are present. For this paper, we prefer to have a clean
definition of disc stars which is the main subject of this
work. In a forthcoming paper,
we will present a detailed study of the properties of the spheroidal
components.

\begin{table*} 
\begin{small}
\caption{Main properties of discs and spheroids obtained for the simulated galaxies:
disc ($M_{\rm d}$) and spheroid ($M_{\rm s}$) masses,  cold gas mass within twice
the optical radius ($M_{\rm cold}$), ratio between disc
 and dark matter masses ($f_{\rm d,DM}$), disc-to-total mass ratio, half-mass
formation times of discs ($\tau_{\rm d}$) and spheroids  ($\tau_{\rm s}$),
half-mass disc radius ($r_{\rm d}$) and half-mass height of disc stars ($h_{\rm d}$).}
\vspace{0.1cm}
\label{table_disks}
\begin{center}
\begin{tabular}{lccccccccc}
\hline
Galaxy  & $M_{\rm d}$ &  $M_{\rm s}$ & $M_{\rm cold}$& $f_{\rm d,DM}$ & D/T &$\tau_{\rm d}$  & $\tau_{\rm s}$  &  $r_{\rm d}$ &  $h_{\rm d}$\\
        & [$10^{10}$M$_\odot$] &   [$10^{10}$M$_\odot$] &   [$10^{10}$M$_\odot$] & &  & [Gyr] &[Gyr]  &  [kpc] &  [kpc]\\\hline

Aq-A-5     & 0.55  &      8.45 &0.151  &0.004 &0.061&  7.63     &   1.73    &    21.2   &    0.42\\

Aq-B-5    &  0.34 &       3.62 & 0.033& 0.005   & 0.087&  8.51 &        3.38 &       24.0   &   0.14\\

Aq-C-5   &    2.30&    8.48 &0.118 &  0.016  &0.213&     4.47 &        1.29 &         12.2&        0.81\\

Aq-D-5& 1.60 &        6.29& 0.001 &0.012  &  0.204  &      4.92&         2.19 &        11.0 &    2.95\\

Aq-E-5 &1.12 &  7.10 & 0.044& 0.012 & 0.136 &   2.90 &     2.23&   12.9&   1.89\\

Aq-F-5 &  -     &    7.70 &0.009&    - & - &  -     & 2.89 & -&-\\

Aq-G-5  & 1.03&    3.37 & 0.060& 0.017  &0.234&   4.63&   1.98&   10.82   & 1.85\\

Aq-H-5&  0.29&     6.19& 0.011&  0.004&0.044&    3.66&     1.56   &10.56   &     5.14\\\hline

\end{tabular}
\end{center}
\end{small}
\end{table*}

\begin{figure}
{\includegraphics[width=42mm]{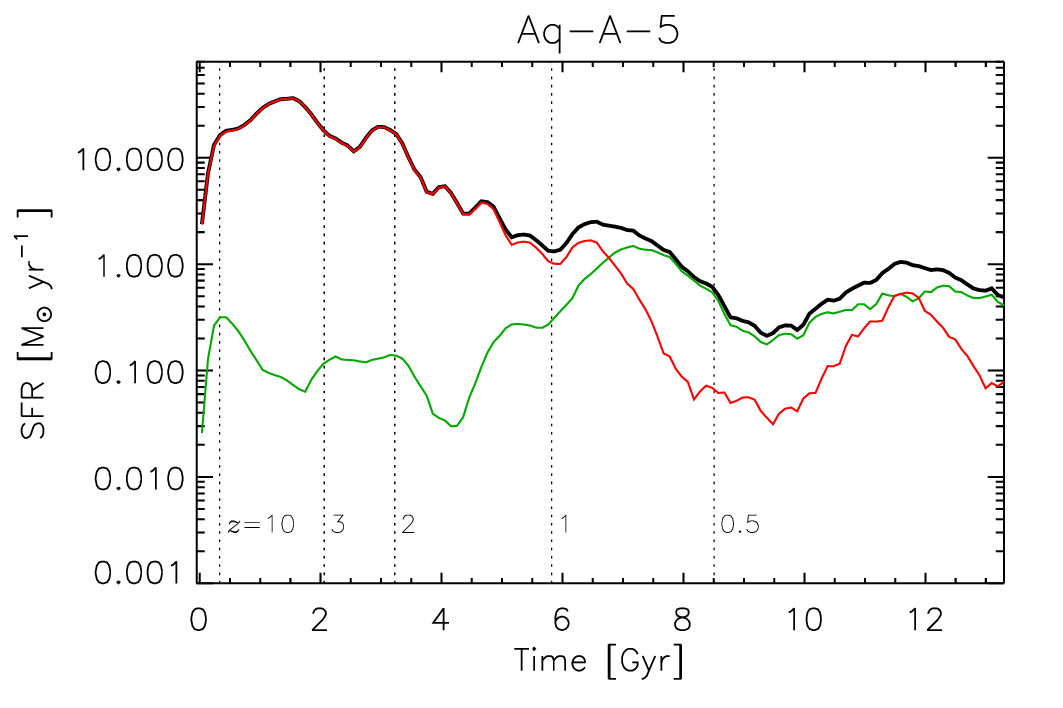}\includegraphics[width=42mm]{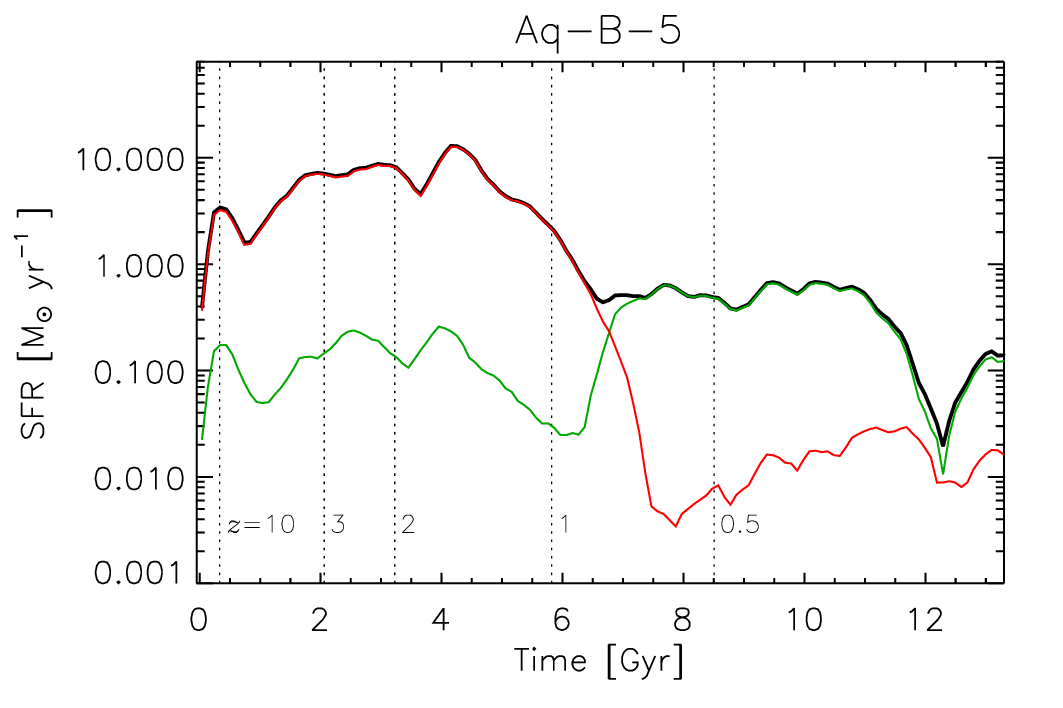}}
{\includegraphics[width=42mm]{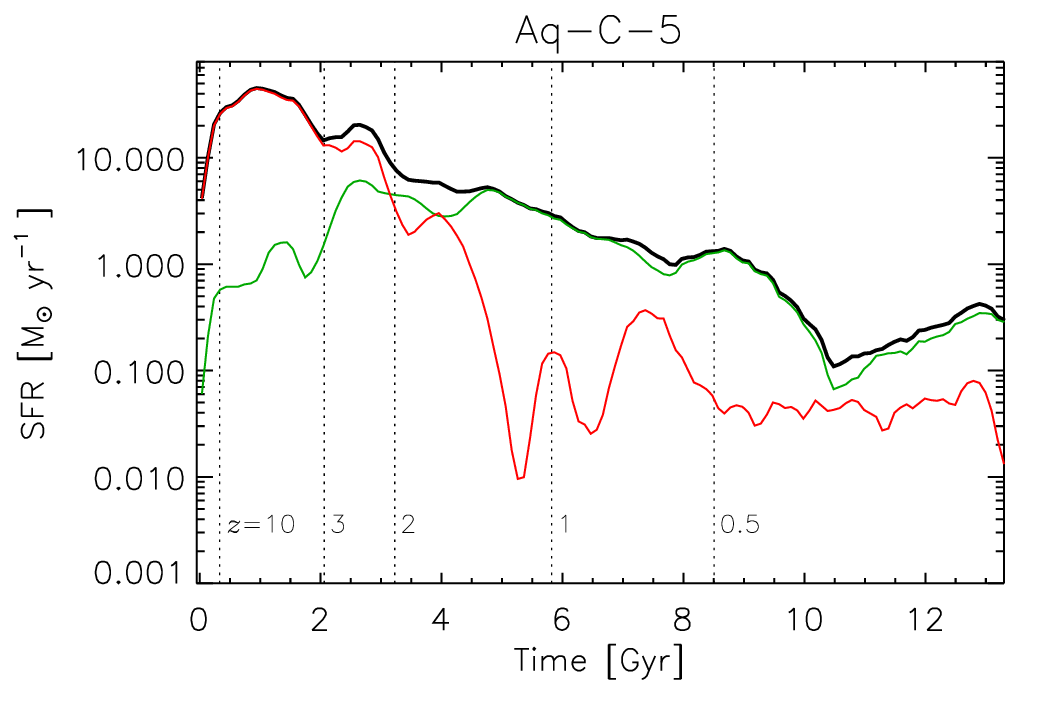}\includegraphics[width=42mm]{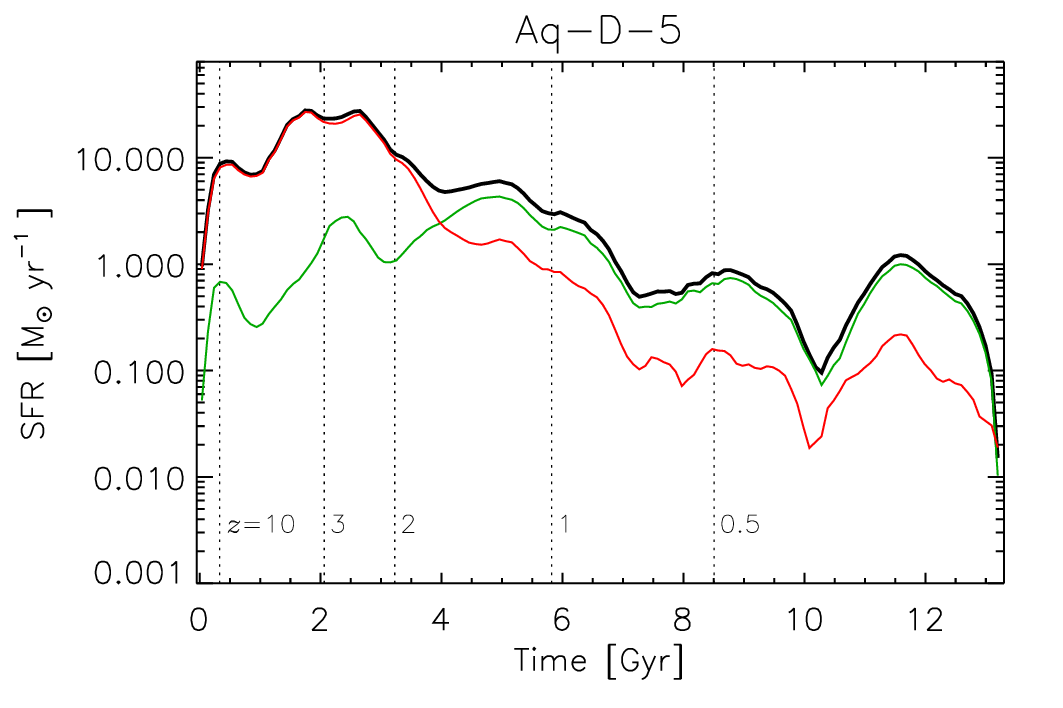}}
{\includegraphics[width=42mm]{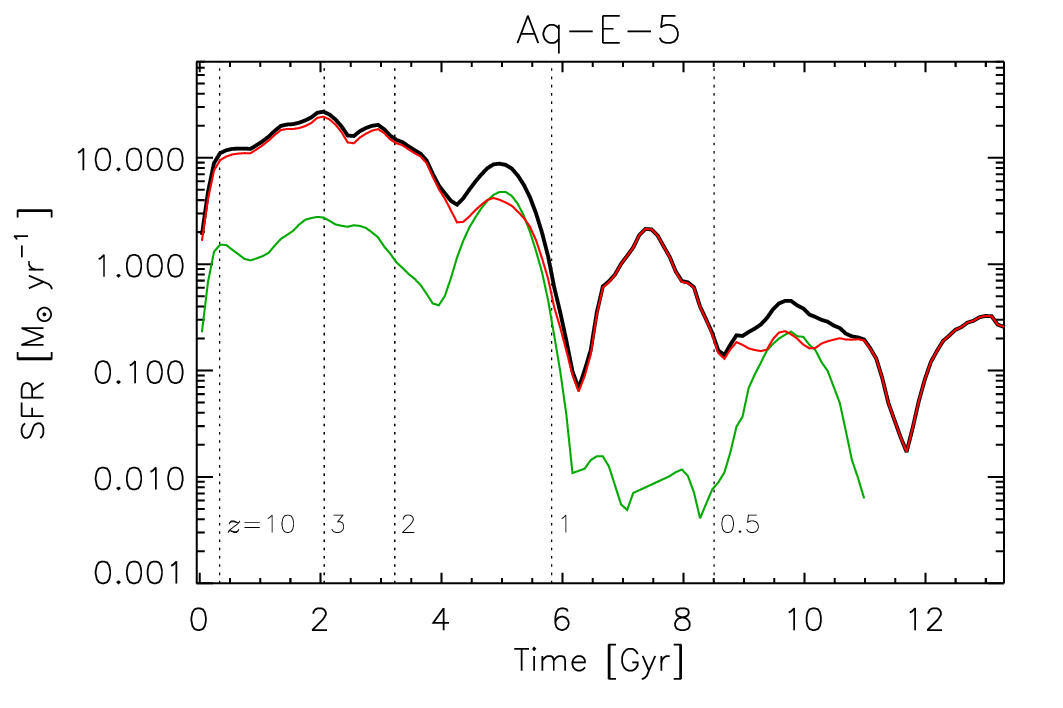}\includegraphics[width=42mm]{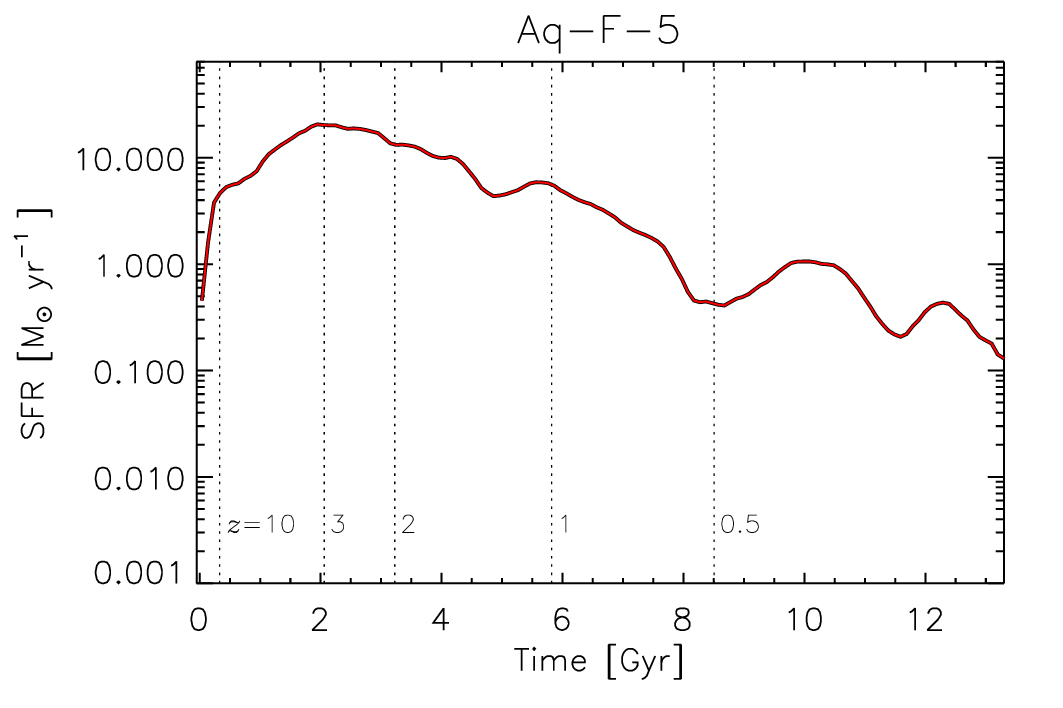}}
{\includegraphics[width=42mm]{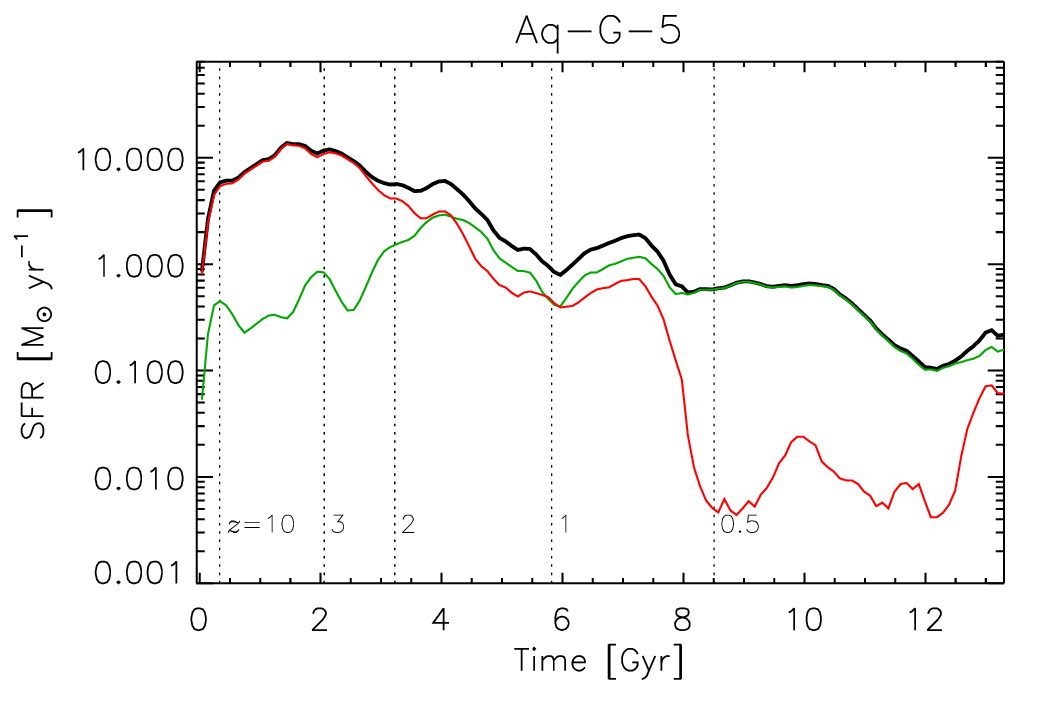}\includegraphics[width=42mm]{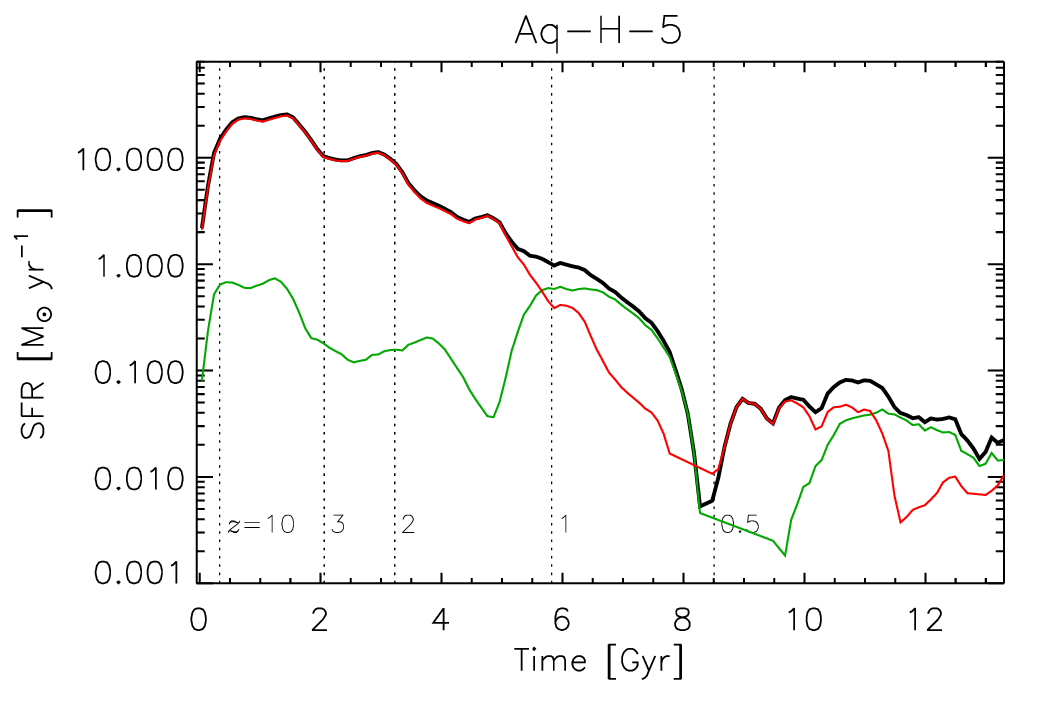}}
\caption{Star formation rates for our eight simulations. Black lines show the total SFRs, while
green and red lines correspond to the SFRs of the discs and the spheroids, respectively.}
\label{fig_sfrs}
\end{figure}

By using the disc-spheroid decomposition described
above, we can study 
how the different components of our simulated galaxies formed,
and how their stellar mass was assembled.
In Fig.~\ref{fig_sfrs}, we show the SFRs for
our set of simulations.
All of them show a similar behaviour, with
an early starburst followed  
by a more quiescent star formation phase at recent times.
The bursty shape found for the SFRs is characteristic
of our SN feedback model (S06; S08). 
We also show  the SFRs constructed separately from disc (green lines)
and  spheroid (red lines) stars.
Clearly, both components have different typical time-scales
of star formation; stars formed at early times contribute
to the formation of spheroids, while recent star formation
leads to the formation of discs.
This can be quantified through the half-mass
formation times of the two components ($\tau_{\rm d}$ and $\tau_{\rm s}$), 
listed in Table~\ref{table_disks}. 
All spheroids are formed very early on, and they present
typical mean formation times of the order of $\tau_{\rm s}\sim 1-3$ Gyr.
In contrast, discs have younger stellar populations, 
with   $\tau_{\rm d}\gtrsim 3-4$ Gyr in all cases.

In Table~\ref{table_disks}, we show the $z=0$ 
masses of the discs and spheroids  for the simulated
galaxies.
From these values, we can see that discs are
substantially less massive than spheroids in all cases (see below).
This  suggests the need to prevent the formation
of massive spheroids at early times: this would
have the double effect of having less massive
spheroids and leaving more gas available for
star formation at recent times.
It is striking that the remaining cold gas fractions in all our
galaxies are very much smaller than in typical spiral galaxies. If more
cold gas were left over, there would be 
a higher SFR at recent times,
and later formation
times for our discs, more similar to observed
spirals
(MacArthur et al. 2004). 
Possible solutions intended to prevent the formation
of spheroids at early times and to retain more gas for later star formation in discs
will be investigated in a separate
work.

Our simulated galaxies have virial masses between
 $7\times 10^{11}$
and $1.6\times 10^{12}$ M$_\odot$.
In order to quantify the  prominence of the discs in each case,
we have calculated the disc-to-total mass
ratio D/T$\equiv M_{\rm d}/(M_{\rm d}+M_{\rm s})$, and
the parameter $f_{\rm d,DM}$, defined as the ratio between disc mass and
dark matter mass within the virial radius.
The 
$f_{\rm d,DM}$ and D/T values obtained for our galaxies
are shown  in  Table~\ref{table_disks}.
Simulations Aq-C-5, Aq-D-5, Aq-E-5 and Aq-G-5
have the highest  $f_{\rm d,DM}$ and D/T ratios, 
consistent with our results on their $\epsilon$
distributions.
In contrast, simulations with small discs
are reflected in very small values of $f_{\rm d,DM}$ and D/T.

\begin{figure}
\includegraphics[width=80mm]{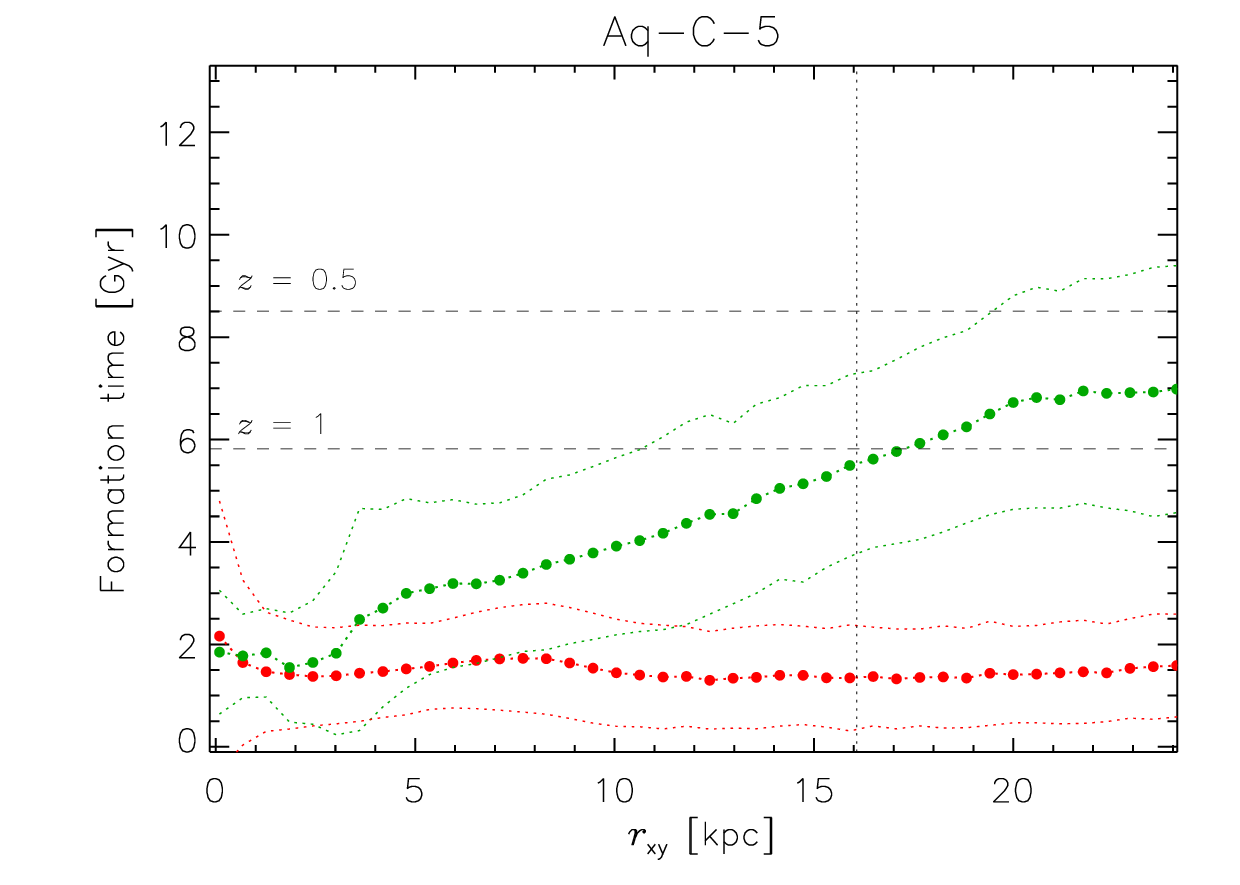}
\caption{Mean formation time of disc (green lines) and spheroid (red lines) 
stars as a function of distance from the rotation axis, for simulation Aq-C-5.
Dotted lines represent the $\pm\sigma$ scatter around the mean.
The vertical line indicates the position of the optical radius of the
galaxy, and horizontal lines have been drawn to indicate the times
corresponding to $z=1$ and $z=0.5$. }
\label{fig_inside_out}
\end{figure}

The D/T ratios obtained for our simulated galaxies, even for those
with well-formed discs, are low in  comparison
with those of real spirals (e.g. Graham \& Worley 2008). 
For comparison, the Milky Way has
a  disc-to-total ratio of $0.95$.
This indicates that, although our
code is able to generate extended galaxy discs and does not suffer
from catastrophic loss of angular momentum, the discs
obtained from this  ``typical'' set of initial conditions are 
much less massive than typical  real discs.
This is probably due to  overly strong feedback
which expels a significant amount of baryons
from the systems and 
removes the new material for disc star formation.
All systems have baryon fractions
of $\lesssim 0.10$ (Table~\ref{simulations_table}).  As discussed above, our results
suggest the need to investigate in more detail the
formation of stars  at early times which
leads to the spheroidal components and the retention of cold disc
gas to fuel  recent
star formation. The low SFRs
we find in all our galaxies are a consequence of the
 very small amount of cold
gas at low redshift, as indicated 
in Table~\ref{table_disks}.

We note, however, that a correct comparison with observations
 is not trivial.
In particular,  D/T ratios calculated from
the simulations refer to stellar mass, not
luminosity, and therefore do not consider
the age of the stellar populations. Dust and inclination
effects are not taken into account either.
Also, our
selection of disc stars based on their kinematics may
exclude  bars, for example, which are labeled as part of the spheroids.
As a result, the D/T ratios listed in Table~\ref{table_disks}
are probably underestimates of the mass of the geometrically thin component.

 We note that although improved numerical resolution might help
to produce more realistic and thinner discs,  the failure
to simulate the formation of late-type galaxies will
not be solved by an increase in resolution. As mentioned
above, we have tested the effects of resolution
by running two lower resolution versions
of Aq-E-5 (Table~\ref{simulations_table}).
In the case of Aq-E-6, we have also performed a full analysis
of the final baryonic systems and find that the
properties of the resulting discs and spheroids
are consistent within a few per cent.
This analysis suggests that the conclusions drawn
from our level $5$ resolution simulations are robust.

We have also calculated  radial and
vertical scale lengths for the  discs in our simulations.
Results are listed in Table~\ref{table_disks}, where
we show the estimated  half-mass radius ($r_{\rm d}$)
and half-mass height ($h_{\rm d}$) for our discs. 
The radial extent of discs varies significantly
between the simulations, but those with substantial discs
all have similar size, $r_{\rm d}\sim 12$ kpc. Their
thicknesses are more dissimilar,  ranging
over $0.8-3$ kpc.

We find that discs typically grow from the inside-out.
As an example, in Fig.~\ref{fig_inside_out} 
we show the mean formation time of disc stars as a function of projected
radius (green line) for simulation
Aq-C-5. We also include  results for the spheroidal
component (red line).
Dotted lines represent the $\pm\sigma$ scatter
around the mean and dashed lines have been plotted
at the times corresponding to $z=1$ and $z=0.5$.
The vertical dotted line indicates the position of the optical
radius.
We find a clear trend between formation
time and radius, indicating that the disc has
grown from the inside out. This behaviour is
common to all discs regardless of their size, except for
Aq-E-5 where we find no trend between formation
time  and radius. Note that Aq-E-5 is the only galaxy in our
sample where the spheroidal component has a net rotation.
Even for those galaxies where the disc component is
very small at $z=0$ (such as Aq-A-5, Aq-B-5 and Aq-H-5),
we find that discs formed from the inside-out.
One interesting feature is that, in
Aq-A-5 and Aq-B-5, the discs have very large
half-mass radii $r_{\rm d}\gtrsim 21$ kpc 
and mean formation times $\tau_{\rm d}\sim 7.6$ Gyr
(Table~\ref{table_disks}). These young discs
 are located farther from the centre,
in comparison to the rest of  our sample, even
those with higher D/T ratios.

The red line in Fig.~\ref{fig_inside_out}  shows 
similar data
for the spheroidal component. In this case, we find
that the stars are all old, regardless
of their distance from the centre of the galaxy.
In all our galaxies (independent of the presence
of a significant disc at $z=0$), the spheroidal components
are similar in terms of stellar age and spatial distribution. 
This suggests that the structure of 
spheroids is not very strongly affected by the presence of discs.

\section{Disc formation and halo properties}\label{halo_prop}

So far, we have analysed the 
properties of discs and spheroids  in our
simulated galaxies. 
In this section, we look for hints to help us understand
the relation between the morphology of galaxies and the properties
of the  dark matter haloes in which  they reside.

\begin{figure*}
{\includegraphics[width=70mm]{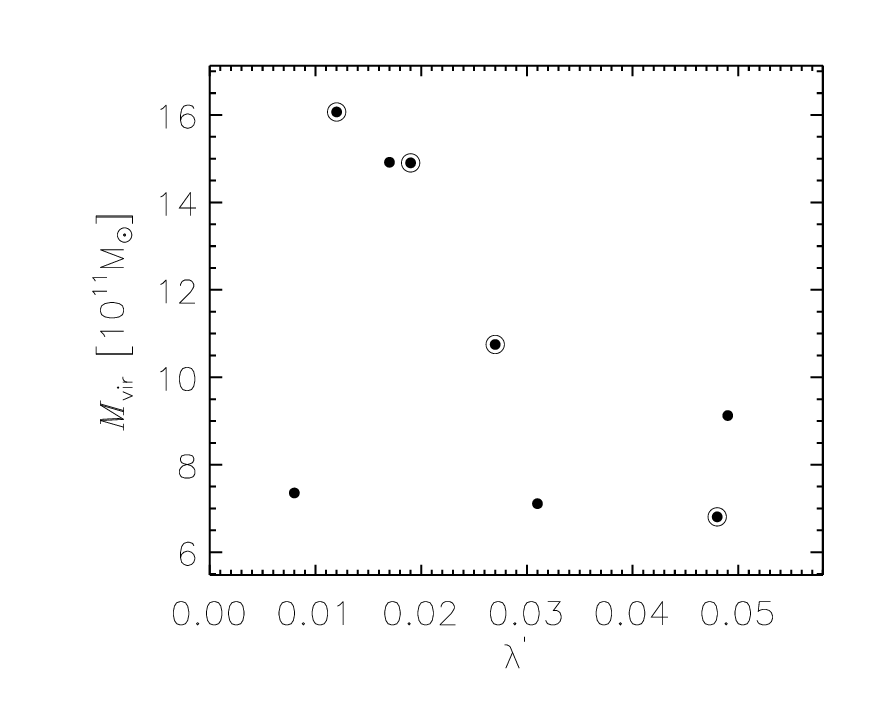}\includegraphics[width=70mm]{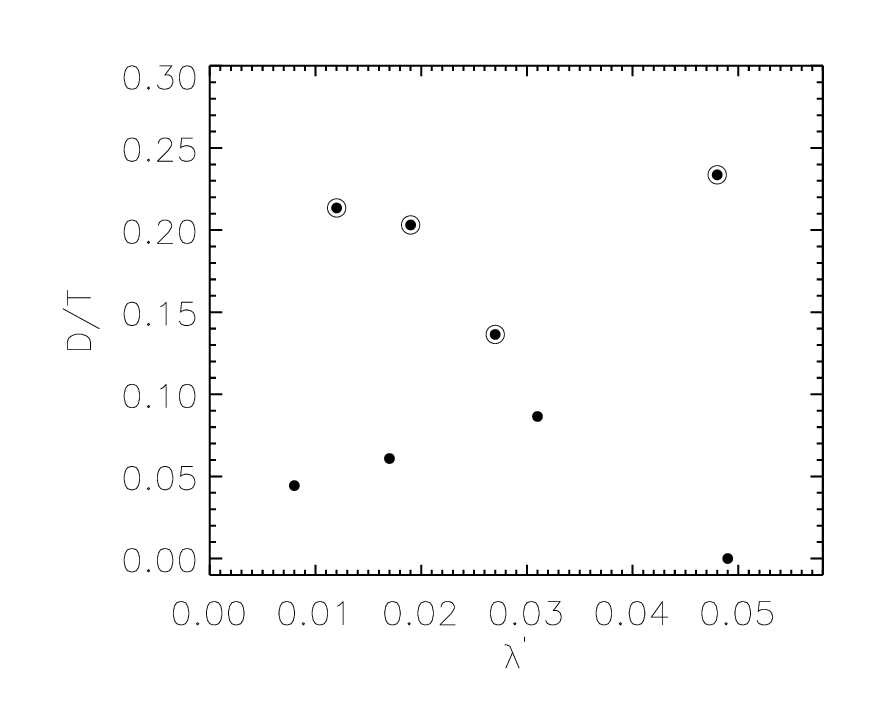}}
\caption{Virial mass  (left-hand panel)  and
disc-to-total mass ratio (right-hand panel) for our simulated galaxies
as a function of the spin parameter of their parent haloes. Encircled symbols
correspond to galaxies with well-formed disc components at $z=0$
(Aq-C-5, Aq-D-5, Aq-E-5 and Aq-G-5).}
\label{fig_spinparam}
\end{figure*}

First, we have investigated possible links between
the presence of discs at $z=0$ and the spin parameters
$\lambda '$ of the haloes, listed in Table~\ref{simulations_table}.
$\lambda '$ has been calculated
using Eq.~(5) of  Bullock et al. (2001) at the virial radius.
In the left-hand panel of
Fig.~\ref{fig_spinparam}, we show the spin parameter of our
haloes as a function of their virial masses.
Spin parameters range between $0.008$
and $0.05$; and in this sample  more massive haloes 
have lower spin parameters.
Encircled symbols correspond to Aq-C-5, Aq-D-5, Aq-E-5 and Aq-G-5, the
haloes where substantial disc components formed.
We find no correlation between the spin parameter
and the presence of significant discs at the present time.
The right-hand panel of Fig.~\ref{fig_spinparam} shows
the disc-to-total mass ratio of our galaxies
as a function of their halo spin parameter.
No trend  
is detected; in these simulations, discs have formed in haloes with spin parameters
as high as $\lambda `\sim 0.05$ and as low as
$\lambda `\sim0.01$.
Galaxies where discs are unimportant
span the same range of spin parameters.

\begin{figure}
\includegraphics[width=80mm]{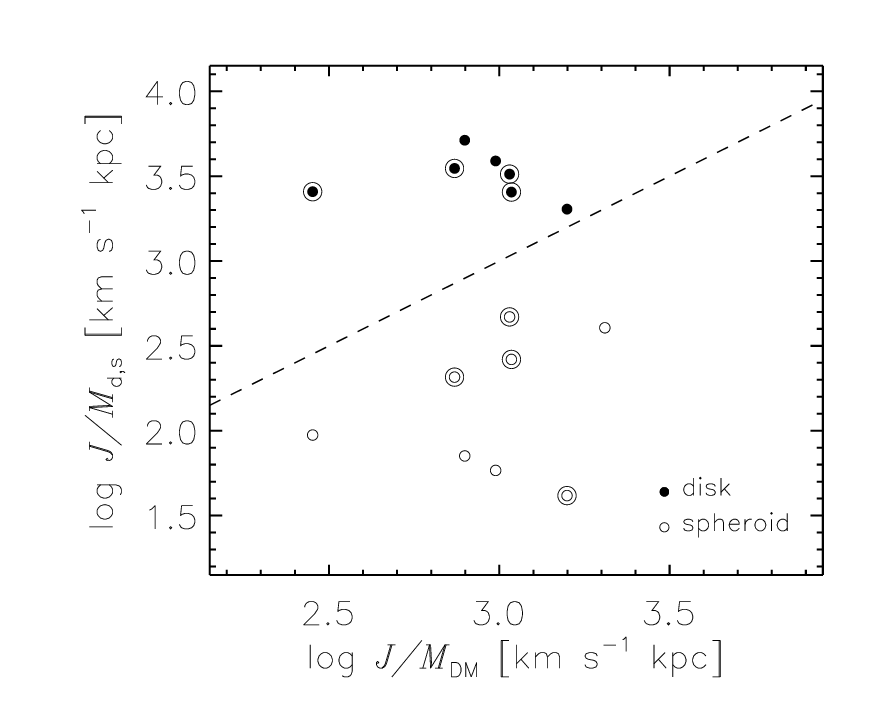}
\caption{ Specific angular momentum of the disc (filled symbols)
and spheroidal (open symbols) components of simulated galaxies,
as a function of specific angular momentum of the dark matter components.
Encircled symbols
correspond to galaxies with well-formed disc components at $z=0$
(Aq-C-5, Aq-D-5, Aq-E-5 and Aq-G-5). The dashed line indicates
the identity relation. }
\label{fig_jspec}
\end{figure}

In  Fig.~\ref{fig_jspec}, we show the
specific angular momentum of the disc and spheroidal components
as a function of the specific angular momentum of their dark
matter halo. As above, encircled symbols
correspond to those haloes where well-formed discs are
present at $z=0$. The dashed line indicates the identity  relation.
We find that spheroids have  lower specific angular momenta
than their dark matter haloes, as expected, but
discs have high specific angular momenta, up to
an order of magnitude larger than
those of their dark matter haloes. 
We find this behaviour for all discs, regardless of their
mass at the present time and of the spin parameter of their parent halo.

\section{Discs at high redshift and disc survival}\label{disk_survival}

We find that all simulated galaxies, regardless of their
D/T ratios at $z=0$, were able to form discs at some
time 
during their evolution.
In order to quantify the relative importance
of discs as a function of time, we have calculated
the disc-to-total mass ratio at a series of redshifts for our
simulated galaxies.
In this case, we have not used the
disc-spheroid decomposition described
in  Section~\ref{results}; this method 
is quite time-consuming and requires
inspection of the individual $\epsilon$ and density distributions.
Instead, we have calculated  D/T  at each
time assuming that the mass of the disc
is given by the total mass of
$\epsilon\ge 0.5$ stars. We find that
this estimation gives slightly higher D/T ratios
at $z=0$ than those obtained with
the disc-spheroid decomposition of Section~\ref{results}.
Because of this,  we have shifted down  the D/T ratios
obtained in this way by a fixed factor $f_{\rm s}$ 
($f_{\rm s}$ is different depending on the galaxy and
it is  independent of redshift),
in order to recover, at $z=0$,  the $z=0$ D/T ratios of Table~\ref{table_disks}.
The $f_{\rm s}$ values are $\lesssim 0.1$ in all cases, except
for Aq-E-5, where $f_{\rm s}=0.3$.
In particular, we find that
seven out of our eight simulated galaxies
had significant disc components at $z\gtrsim 2$,
and six of them at $z\sim 1$.

\begin{figure*}
\includegraphics[width=185mm]{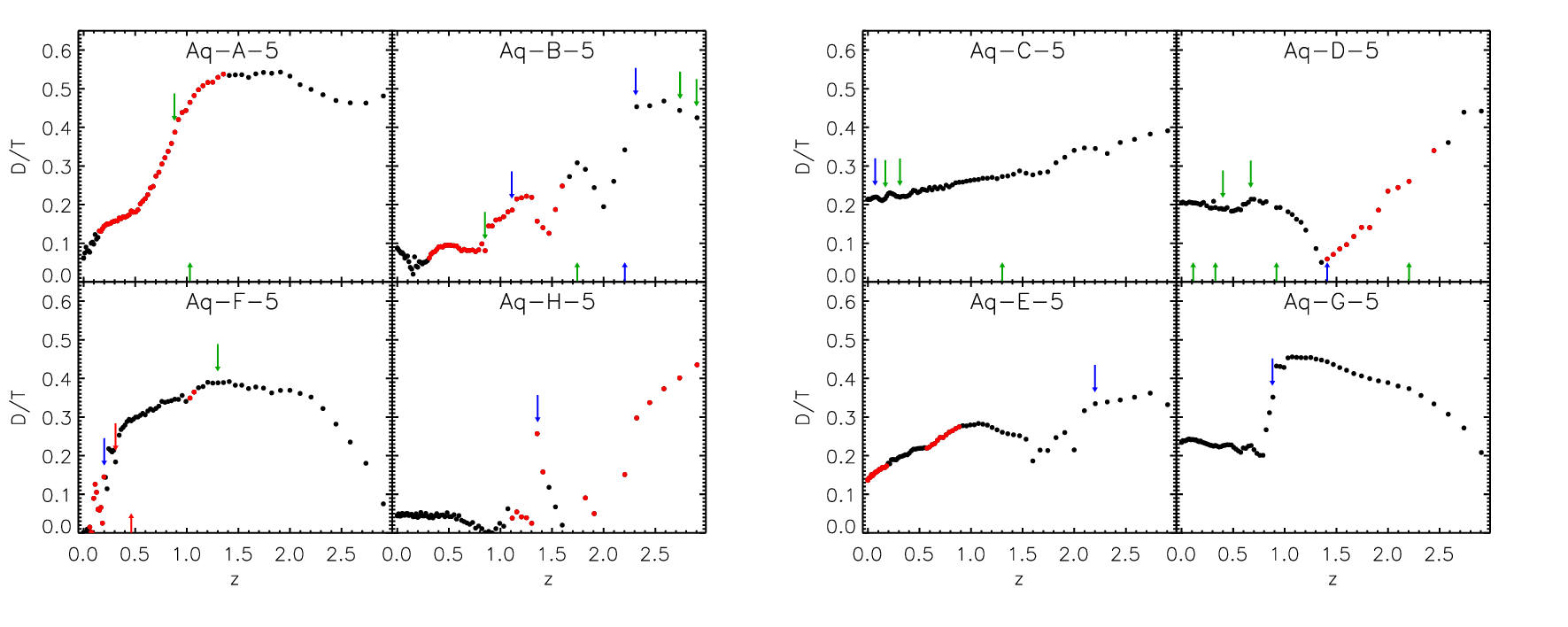}
\caption{Disc-to-total mass ratio as a function of redshift for simulated
galaxies. We have divided the plot in such a way that galaxies with
little or no disc components at $z=0$ are plotted in the left-hand
panel, while galaxies with more prominent discs at the present time reside
in the right-hand one. The arrows indicate the entrance of satellites
either to the virial radius (upward arrows) or to the central (comoving) 
$27$ kpc (downward arrows). 
Arrows are colour-coded according
to the merger ratio $f\equiv M_{\rm sat}/M_{\rm cen}$:
 red, blue and green colours correspond to $f>0.3$,
$0.1<f\le 0.3$, and $0.02<f\le 0.1$, respectively.
Red points indicate periods of strong misalignment between
the cold gaseous and stellar discs: cos $\beta$$<0.5$, 
see Fig.~\ref{fig_cos_beta}.}
\label{fig_dt}
\end{figure*}

Fig.~\ref{fig_dt}  shows
the evolution with time of the D/T ratios for our eight simulated galaxies.
We have  divided the plot in such a way  that galaxies with
little or no disc components at $z=0$ are in the left-hand
panel, while galaxies with more prominent discs at the present time
are in the right-hand one.
Clearly, the relative importance of discs and
spheroids at the present time has been reached 
in different ways. 
Let us first concentrate on those galaxies
with little disc component  at $z=0$ (left-hand
panel).
It is clear that these  galaxies have discs  at higher redshifts;
discs did even dominate the stellar mass, such as  in  Aq-A-5.
Later on, however, these discs  transferred most of their mass
to the spheroids, although in different ways.
Aq-A-5 had a large disc which
was quite stable until $z\sim 1$; after
 this time, its D/T ratio  decreases steadily.
A similar behaviour is found for Aq-F-5,
the only galaxy of our sample with no disc component at $z=0$.
Aq-B-5  undergoes a series
of sharp changes in its D/T ratio, and a new disc  started
to grow near $z=0$. Finally, Aq-H-5
had a significant disc only at $z\gtrsim 2$ which
dissolved by $z\sim 1$; after this
time, a new disc  formed, although
it was not  able to grow enough to produce
a substantial component by $z=0$.

The galaxies plotted in the right-hand panel
of Fig.~\ref{fig_dt} 
did produce important discs by $z=0$.
Here, we also detect differences in the way discs formed.
In Aq-C-5 and Aq-E-5, discs were present even at
$z=3$  and they remain stable until $z=0$.
Aq-D-5 had a significant disc at $z=3$ that
dissolved by $z\sim 1.5$, but
a new disc formed between $z=1.5$ and $z=1$ which
survived until the present time. Aq-G-5 
had  quite a massive disc at high redshift with a maximum
D/T ratio of about 0.5 at $z\sim 1$; later, however,  the disc starts
losing mass to the spheroid to reach $z=0$ with D/T $\approx 0.2$.

Our findings suggest that discs are quite common
at high redshift, but they often do not survive
until the present time. 
The critical question is clearly  what determines the survival
or destruction of discs.
It is generally thought that mergers and disc instabilities are
the principal processes 
contributing  to a change in galaxy morphology.
Here, we investigate some features that 
affect the evolution of discs and spheroids in
our simulated galaxies.
On one hand, we investigate the effects
of mergers on the disc components.
On the other hand, we study whether misalignment between
the gaseous and stellar discs can trigger instabilities
which 
transfer  mass from the discs to the spheroids.

\begin{figure*}
\includegraphics[width=175mm]{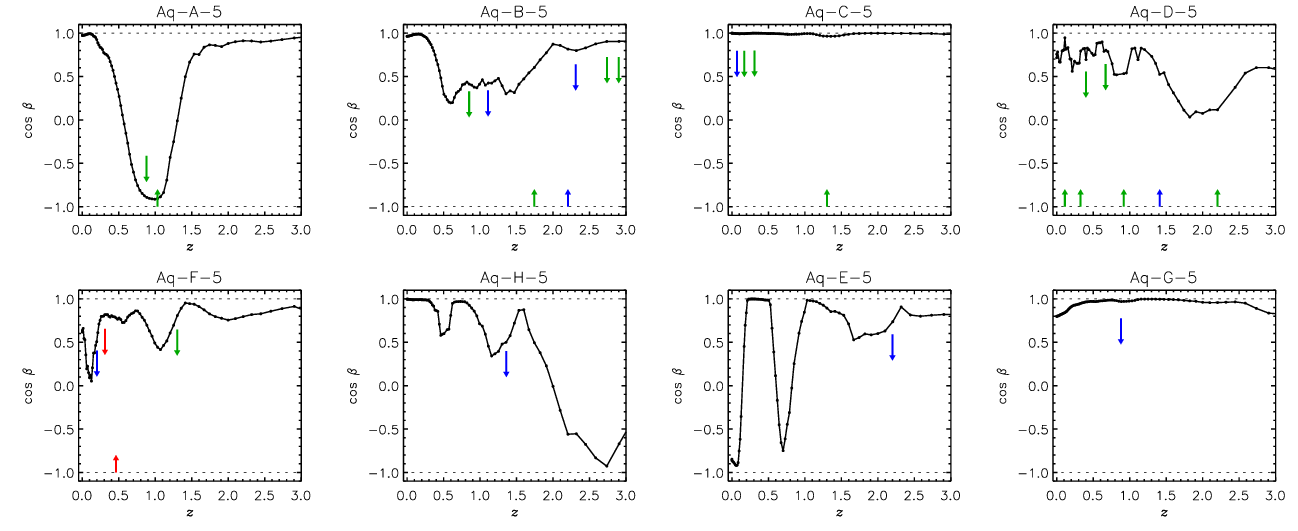}
\caption{Cosine of the angle $\beta$ between the angular momentum
vectors of the cold gas and stars in the inner $27$ comoving
kpc as a function of redshift.
 The plot has been organized in such a way that galaxies with
little or no disc component at $z=0$ are plotted on the left
panels, while galaxies with more prominent discs at the present time are on
 the right.
Arrows indicate the entrance of satellites
either within the virial radius (upward arrows) or within the central (comoving) 
$27$ kpc (downward arrows). Arrows are colour-coded according
to the merger ratio $f\equiv M_{\rm sat}/M_{\rm cen}$:
 red, blue and green colours correspond to $f>0.3$,
$0.1<f\le 0.3$, and $0.02<f\le 0.1$, respectively.
}
\label{fig_cos_beta}
\end{figure*}

Major mergers are known to destroy
discs and to produce spheroidal remnants.
In order to illustrate this effect, 
we have included in Fig.~\ref{fig_dt}  two sets of arrows 
which show when satellites enter the 
inner $27$ comoving
kpc (downward arrows), or the virial radius (upward arrows).
The arrows are colour-coded according to $f\equiv M_{\rm sat}/M_{\rm cen}$
where  $M_{\rm sat}$ and $M_{\rm cen}$ are the
stellar masses of the satellite and of the central
object, respectively. We include  only those events
where $f>0.02$. Red arrows indicate events with $f>0.3$,
blue arrows correspond to $0.1<f\le 0.3$, and green ones
to  $0.02<f\le 0.1$\footnote{In terms of 
total masses, we find lower values for the relative masses
of the satellite and the central objects, this indicates that the
dark matter is more strongly stripped than the stars.}.
We define major mergers as those with $f>0.3$  (red arrows).  
With this definition, only Aq-F-5 experiences a major merger
($f=0.4$) which happens at $z\sim 0.5$, and  completely destroys the disc.
This galaxy is the only one in our sample that, 
at $z=0$, has no disc component at all.

Mergers where $0.10<f\le 0.30$ 
can also produce  important effects on the
baryonic mass distributions. We find that, in
some cases, such events
coincide with a decrease in the D/T ratios, as
is most clearly seen in the case of
Aq-G-5. In this galaxy, the D/T ratio
drops rapidly from $\sim 0.45$ to $\sim 0.2$
after a merger event (with $f=0.17$);
however, the disc  is able to survive until $z=0$. 
Smaller satellites ($f<0.1$) do not seem to significantly
affect the mass distributions.
In general, we find that the effects
of mergers with similar $f$ values can be different
for different galaxies, presumably this partly depends
on the orbit of the infalling satellite. 
These results suggest that satellites
may significantly affect galaxy evolution, 
depending on both the relative mass of the satellite
and central systems and the satellite's orbit.

We now investigate another process which can lead to the destruction
or shrinking of discs, this is the misalignment between
the angular momentum of infalling cold gas and that of the stars
in the centre of galactic haloes.
Infall of intergalactic or interstellar gas, as well as of satellites,
can induce this behaviour.
We find that misalignment between the gaseous and stellar
discs is quite a common feature in our simulations,
and can significantly contribute to the transfer of mass
from the discs to the spheroids.
In  Fig.~\ref{fig_cos_beta}, we show the cosine
of  $\beta$, the angle between the angular momentum vector
of the cold gas and that of the stars  in the  central
region of the galaxies,
as a function of redshift\footnote{ We have used an
arbitrary threshold radius of $20\ h^{-1}=27$ comoving kpc.
Provided most baryons are within this region, our results
are not sensitive to this choice.}
The first two columns show  results for Aq-A-5, Aq-B-5, Aq-F-5
and Aq-H-5, which have big spheroids and little discs at $z=0$;
whereas  the last two columns  correspond to Aq-C-5,  Aq-D-5, Aq-E-5 and
Aq-G-5, galaxies with more prominent discs.
We find clear differences in the evolution of $\beta$ for
these two groups of galaxies.
Those in the second group, which have significant disc
components at $z=0$, typically have cos $\beta$$\sim 1$ since $z=3$
(except for Aq-D-5 where, around $z\sim 2$, the
angular momenta of the cold disc and of the stars are almost 
perpendicular).  (In the case of Aq-E-5, we detect 
two periods of misalignment corresponding to infalling
cold clumps.)
On the contrary, those
galaxies in the first group (with little disc at $z=0$)
show significant and rapid changes in cos $\beta$ with redshift.
This indicates that the  cold gas and stars
are not coplanar.  Note that a second stellar disc,
not aligned with the already existing one, may be formed
from the cold gas, provided it  is dense enough.
In some cases, the cold gas and stellar components are even
roughly counter-rotating, such as in
Aq-A-5 and Aq-B-5  at $z\sim 0.5$, and in Aq-H-5 at $z\gtrsim 2$.

The interaction between two misaligned discs
destabilizes the systems, triggering a redistribution
of mass  from the discs to the spheroids.
This can be seen in Fig.~\ref{fig_dt}, where we have drawn in red
those points  where cos $\beta$$\le 0.5$.
Clearly, a high misalignment between the gaseous and stellar
discs is always accompanied by a decrease in the D/T ratio, 
indicating that mass is being transferred from the  disc
to the spheroid. Note that, at high redshift, the
decrease in D/T may also indicate that spheroids are
getting more massive, not at the expense of the disc. 
However, for $z\lesssim 2$ there 
are almost no new spheroid stars, as can be inferred 
from the star formation rates of spheroid stars
(Fig.~\ref{fig_sfrs}) and from the spheroid half-mass formation
times $\tau_{\rm s}$ (Table~\ref{table_disks}),
and therefore D/T can only decrease if stellar
mass from the discs is transferred to the spheroids.

Our results show that both major mergers and disc instability,
where mass is transferred from  discs to  spheroids,
contribute to the destruction and/or shrinking of discs. 
Taking into account the results of the previous
section, we conclude that the presence of discs at $z=0$ is more
likely determined by the particular formation and merger
histories of galaxies, rather than by the spin parameter
of their parent haloes. The absence of major mergers
seems to be a necessary condition for discs to form, but
not to be sufficient.

\section{Conclusions}
\label{conclu}

We have used a series of simulations to study the formation
of galaxies in  a $\Lambda$CDM cosmology.
Our simulations have enough resolution to follow the internal
structure of galaxies and, at the same time, to keep track
of larger scale processes such as mergers, infall and tidal stripping. 
We have simulated the evolution of eight galaxies with
virial masses in the range $7-16\times 10^{11}$M$_\odot$.
These were selected purely on the basis of their mass
  ($\sim 10^{12}$ M$_\odot$) and in isolation (no companion exceeding half of their mass
within $1.4$ Mpc), 
with no additional restriction on spin parameter or  merger history.
The only condition  was the absence of  close
massive companions at $z=0$.
As a result, our sample, although still small, should be representative 
for the formation of isolated galaxies similar to the Milky Way.

Dark matter only simulations of these haloes 
with much higher resolution than those studied here have been
carried out for the
Aquarius Project of the Virgo Consortium (Springel et al. 2008). 
In this paper, we have included gas dynamics using a specialized extension
of the code {\small GADGET-3} which includes star formation and
supernova feedback (both from Type II and from Type Ia SNe), metal-dependent
cooling, a multiphase treatment of the gas component, and photoheating by the
UV background (S05; S06).
This hydrodynamical code includes no  scale-dependent
parameters, and is among the most detailed and
least ad hoc attempts to describe the interstellar medium  physics
and  SN feedback in galaxy formation.
 One of
its main  advantages is that galactic winds are naturally
generated, without the ad hoc prescriptions required in
most alternative
simulation codes.

With the parameters adopted here, none of our eight haloes grows
a $z=0$ disc with more than about $20$ per cent of the total
stellar mass. Four
of the simulated galaxies do have  well-formed
and substantial discs in rotational support, three have dominant
spheroids and little discs, and one has no disc at all. 
The star formation histories of all galaxies are similar, with
an early starburst followed by a more quiescent phase
and little star formation at $z=0$. 
Old stars populate the spheroids, while younger stars 
are found in the discs. As a result, typical
half-mass formation times are $\tau_{\rm s}\sim 1-3$ Gyr
for spheroids and  $\tau_{\rm d}\gtrsim 4$ Gyr for
discs.
We find that, regardless of  the presence or absence of a disc,
all spheroids have  similar mass profiles
and stellar age distributions. While all
discs formed from the inside-out, their mass distributions
show greater variety.

Even the most massive  discs formed in these simulations produce
disc-to-total mass ratios of the order of $0.2$, 
much lower than those of late-type spirals or of
the Milky Way. This is primarily
because of the small amount of cold gas available
for star formation at late times. 
Presumably, this  is either due to an excess
of star formation at early times which leaves
the systems with too little gas to make a large disc,
or due to  too efficient feedback
which expels the gas which would otherwise make the disc.
We will investigate this
problem in more detail in a forthcoming paper.

We find  that disc growth in
 dark matter haloes does not depend on spin parameter. We have
significant discs in haloes with spin parameter as low as
$0.01$ and as high as $0.05$, and galaxies with little
or no disc component in haloes with the same range of spin parameters.
We also find  no relation between disc mass fraction and halo
spin parameter.

As expected, our  discs have high specific angular momentum,
even higher than that of their
dark matter haloes. In fact, the discs in our simulations
all have  similar specific angular momenta, independently of their mass.
Spheroids, on the contrary, 
span a wide range of specific angular momentum, although in all cases
their values are smaller than those of the parent dark matter
haloes.
Those spheroids with the highest specific
angular momenta are found
in haloes which have prominent discs at $z=0$.

The variation of disc-to-total ratio with redshift reveals
that all our galaxies were able to form discs at some time during their
evolution. At high redshift,  some discs  are comparable in mass
to their spheroids. However,  discs often
fail to survive until the present time.
We find that both major mergers and misalignment between the
  discs and newly accreted gas  can completely or partially destroy
them. Only one of our galaxies experiences a major
merger which completely  destroys its disc
and, at $z=0$, it only has a spheroid. 
Misalignment between stellar discs
and newly accreted cold gas is more frequent and induces instabilities 
which transfer a significant amount of mass
from the discs to the spheroids. 
Among the galaxies which do have significant
discs at $z=0$,
there are substantial
differences in disc evolution.
 Discs can be stable since $z=3$;
they can be completely destroyed and can regrow; or they can 
lose part of their mass but survive until $z=0$.

Our results suggest that  discs can
form easily in $\Lambda$CDM scenarios, regardless
of the spin parameter of their parent  haloes.
However, their survival  is strongly
affected by
major mergers and by disc instabilities.
It is clear that the particular implementation of star formation
and feedback physics in our current models, while it produces
individual objects which resemble real galaxies, cannot
produce a galaxy population which matches observation. In particular,
the disc mass fractions of our simulated galaxies are
typically much lower than those of real galaxies of
similar stellar mass. We produce nothing which looks
like the Milky Way or other late-type spirals. Apparently,
real galaxies formed fewer stars at early times than
our models and substantially more stars in a disc at
late times.

\section*{Acknowledgments}

We thank the referee for his/her comments that
helped improve the paper.
The simulations were carried out at the 
Computing Centre of the Max-Planck-Society in Garching.
This research was supported by the DFG cluster of 
excellence `Origin
and Structure of the Universe'.
CS thanks A. Jenkins and A. Ludlow for generating
the initial conditions used in this paper.
This work was partially supported by PICT  32342 (2005),
PICT  Max-Planck 245 (2006) of Foncyt and the Secyt-DAAD
bilateral collaboration (2007).

\appendix

\section[]{Disc-spheroid decomposition}\label{ap_A}

In this Appendix, we explain in more detail the
procedure we follow to assign stars to the
disc and the spheroidal components.
In particular, we explain how we construct the
white curves shown in Fig.~\ref{fig_epsilon_vs_r},
which are used to avoid contamination of the disc
by spheroid stars.
Note that contamination will be present
if  the decomposition into the disc and spheroidal
components is
based on $\epsilon$ only. The most clear 
example is Aq-F-5, where $4$ per cent of the stars
have $\epsilon\gtrsim 1$ (compatible with disc
dynamics). However,  this system is a
pure spheroid. All simulations would suffer from this
contamination if no further restriction were
done for disc stars.

As explained in the text, our disc-spheroid decomposition is based on
three conditions: $\epsilon\sim \epsilon_{\rm peak}$,
orbits roughly contained in the disc plane and
disc stars lying above the white lines shown in
Fig.~\ref{fig_epsilon_vs_r}.
The white lines are constructed in the following way. 
(i) For each simulated galaxy, we first draw the lower white line,
 using the following functional form,
 \begin{equation}
y = \epsilon_{\rm s} - C1 - {\rm exp}(-(r_{xy}-C2)/C3) 
\end{equation}
and choosing values for $C1$, $C2$ and $C3$  
 in order to enclose most $\epsilon<\epsilon_{\rm s}$ stars
($> 90$ per cent)
between this line and a horizontal one
at  $\epsilon=\epsilon_{\rm s}$.
(Note that $\epsilon<\epsilon_{\rm s}$ stars can be used as clean tracers
of the spheroid.)
The detailed
shape of the curve is not important, provided
it encloses most of the spheroid stars.
The upper white line is then constructed by reflecting the lower
line symmetrically about
$\epsilon_{\rm s}$. The white lines thus delimit the spatial
region of the velocity dominated spheroidal component.
(ii) As a safety check, we calculate the properties of the
resulting disc and spheroid stars (stellar age, metallicity,
mass profile).
Since the global properties of the disc and spheroidal components
in our simulated galaxies are
distinct and well-defined, contamination is easily detected.
For example, if high $\epsilon$ stars in 
Aq-F-5 are identified as disc stars, they will in any case
be old, metal-poor, and centrally concentrated. These
stars have the same characteristics as the spheroidal
component present in this simulation, and clearly belong to it, rather
than to a disc. Our procedure is successful in avoiding this
contamination.
(iii) If contamination is detected, we 
adjust our white lines, and repeat (ii) until we
get clean samples for the disc and spheroidal components.

Note that this procedure is valid provided the disc
and spheroidal components have distinct and well-defined
properties, which is the case for our simulated galaxies.
We have tried a number of other ways to avoid  contamination
of the discs by spheroid stars, and found this particular one to
be the most satisfactory one 
since it can be successfully applied to all our simulations, including
Aq-E-5, where the decomposition is the hardest.

\end{document}